\documentclass[10pt, twocolumn, floatfix, superscriptaddress, aps, prx]{revtex4-2}
\usepackage{amsmath, amssymb, graphicx, hyperref, xcolor, microtype, zx-calculus}
\usepackage[export]{adjustbox}
\usepackage{braket}
\usepackage{enumitem}

\begin{document}

\title{Quantum many-body scars with tunable entanglement\texorpdfstring{\\}{ }and Hamiltonian inverse design from ZX-calculus}

\author{Marcin Szyniszewski}
\affiliation{Department of Computer Science, University of Oxford, Parks Road, Oxford OX1 3QD, UK}

\begin{abstract}
    Diagrammatic languages such as ZX-calculus provide compact and intuitive descriptions of quantum processes and have become established tools for circuit simplification, verification, and compilation. However, their potential as a framework for constructing many-body states and the Hamiltonians that host them remains largely unexplored. Here, we introduce families of fractal many-body states obtained from ZX-diagrams based on the Sierpi\'nski triangle and Sierpi\'nski carpet. By construction, the underlying graph connectivity imposes atypical subvolume-law minimum-cut upper bounds on entanglement: at most logarithmic scaling for the triangle family and sublinear power-law for the carpet family. By adjusting the free parameters of the ZX-diagram construction, we can tune the actual entanglement scaling: we demonstrate both area-law and logarithmic scaling for the triangle, and both log law and power-law for the carpet. Additionally, their local observables retain fractal-like spatial structure, identifying these states as natural candidates for atypical eigenstates in otherwise thermalizing systems. For the triangle family, we combine parent-Hamiltonian methods, insights from ZX-calculus, and local ZX identities that certify exact annihilation of the target state, producing frustration-free Hamiltonians whose terms admit simple representations in the same diagrammatic language as the states themselves. We then construct a local deformation that produces chaotic level statistics while embedding the fractal ZX state in the bulk of the energy spectrum as an exact quantum many-body scar. Our results demonstrate, through this explicit construction, that ZX-calculus can serve as a framework for Hamiltonian inverse design, in which quantum many-body scars, their local annihilators, and the chaotic Hamiltonians embedding them can be constructed and related through a set of graphical identities.
\end{abstract}

\maketitle

\section{Introduction}

Understanding the thermalization of isolated quantum many-body systems remains a central problem in nonequilibrium quantum physics. In generic nonintegrable interacting models, thermalization is commonly understood through the lens of the eigenstate thermalization hypothesis (ETH), according to which individual finite-energy-density eigenstates give thermal expectation values for local observables~\cite{Deutsch1991, Srednicki1994, Rigol2008, DAlessio2016}. This is closely tied to the concept of quantum chaos, where spectral diagnostics can often be described by random matrix theory~\cite{Bohigas1984, Heller1984}. Important exceptions to this paradigm include integrable systems, many-body localization, Hilbert-space-fragmentation, and systems with special nonthermal eigenstates embedded in otherwise thermal spectra~\cite{Berry1977wfn, Polkovnikov2011, Nandkishore2015, Abanin2019, Khemani2020}. Quantum many-body scars (QMBS) provide one particular example of such weak ergodicity breaking: a small set of atypical, highly-excited eigenstates can coexist with otherwise thermal eigenstates and produce long-lived coherent dynamics~\cite{Turner2018, Turner2018Rydberg, Serbyn2021, Chandran2023, Moudgalya2022}. The modern study of QMBS was stimulated by experiments on Rydberg-atom arrays~\cite{Bernien2017}, and by the subsequent identification of scarred eigenstates in the corresponding constrained PXP model~\cite{Turner2018, Turner2018Rydberg}.

The construction, classification, and stabilization of quantum many-body scar states have become central theoretical challenges, and a variety of complementary mechanisms have now been identified. Projector-embedding constructions insert chosen target states into the bulk of the spectrum of local Hamiltonians~\cite{Shiraishi2017}. Kinetically constrained models and deformations of, e.g., the PXP chain reveal approximate scar towers, forward-scattering structures, and emergent approximate SU(2) dynamics~\cite{Turner2018Rydberg, Choi2019, Bull2019, Ho2019}. Exact scars have also been found in Rydberg-blockaded chains~\cite{Iadecola2020, Lin2019}, AKLT-type models~\cite{Moudgalya2018ent, Moudgalya2018, Moudgalya2020}, spin-1 XY magnets~\cite{Chattopadhyay2020, Schecter2019}, Hubbard models with $\eta$-pairing structure~\cite{Moudgalya2020eta}, systems with spectrum-generating algebras~\cite{Pakrouski2020, Moudgalya2024, Mohapatra2026}, and models with Bell-pair structures~\cite{Langlett2022, Udupa2023, Chiba2024, Mohapatra2025, Ivanov2025, Mukherjee2026, Dooley2026}. Certain scar constructions are particularly notable in that they permit tunable entanglement scaling~\cite{Teretenkov2024, Mukherjee2026}.
\begin{figure}[b]
    \centering
    \includegraphics[width=\columnwidth]{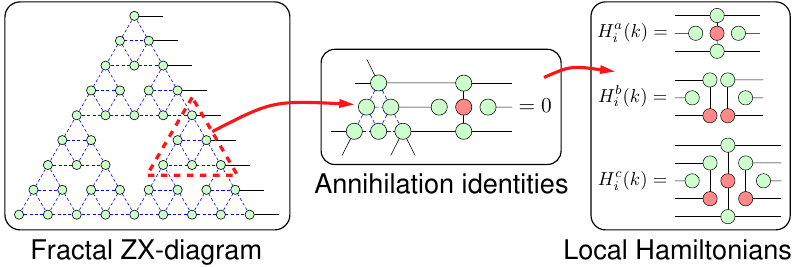}
    \caption{Summary of the methodology: a fractal ZX-diagram state with atypical properties and tunable entanglement is selected as a candidate quantum many-body scar; parent Hamiltonian techniques and local ZX identities are used to find annihilators; associated local Hamiltonians are constructed that host the fractal ZX-diagram as a scar.}
    \label{fig:summary}
\end{figure}
Tensor-network methods have played a particularly important role in this development: exact scarred eigenstates may admit matrix-product-state (MPS) representations, while MPS-based parent-Hamiltonian constructions provide a systematic route to families of scarred Hamiltonians~\cite{Lin2019, Moudgalya2018ent, Moudgalya2018, Moudgalya2020, Chattopadhyay2020, Schecter2019, Shibata2020, ODea2020, Zhang2023}. These results suggest that the structure of constructed nonthermal eigenstates is often more transparent than the structure of the full Hamiltonian, motivating the search for new languages in which both such states and the associated Hamiltonians can be designed directly.

Many exact scar states, particularly low-entanglement constructions, are naturally viewed as low-complexity tensor-network states embedded in a high-complexity part of the spectrum. However, ordinary tensor-network notation is primarily a representation of multilinear algebra, which, although convenient numerically, sometimes obscures the underlying physics. In contrast, the ZX-calculus is a graphical language for qubit linear maps equipped with local rewrite rules~\cite{Coecke2011, Duncan2020}, which often makes algebraic structure and circuit identities transparent, while exposing the underlying quantum processes. Originally developed in categorical quantum mechanics and quantum information theory, it has become a practical tool for quantum-circuit optimization and verification~\cite{Duncan2020, vandeWetering2020, Coecke2017, KissingerWetering2024Book}.
Recently, ZX-calculus and closely related graphical languages have begun to move towards applications in condensed matter physics~\cite{East2022, Wang2025}. ZX-diagrams have been used to represent AKLT states, derive their matrix-product structure, recover edge modes and string order, and construct higher-dimensional topological phases diagrammatically~\cite{East2022}. ZX has also been adapted for SU(2) models~\cite{Wang2025, East2021}, while being able to give new perspectives on equilibrium and dynamical phase transitions~\cite{BuznachAhituv2025, Ha2025}. ZX-based tensor-network methods have also been proposed as diagnostics of topological order~\cite{MasMendoza2025, Lu2025}, as well as facilitating descriptions of gauge theories~\cite{Gorantla2025, Wong2026} and quantum gravity~\cite{Priestley2025}. These developments imply that graphical languages may be useful for constructing and understanding physically interesting quantum many-body systems and their eigenstates.

In this work, we develop such a construction by combining projector-embedding techniques and insights from ZX-calculus, specifically using self-similar ZX-diagrams (see Fig.~\ref{fig:summary}). We introduce two families of many-body states generated by ZX-diagrams based on the Sierpi\'nski triangle and the Sierpi\'nski carpet. The fractality here is not of a real-space lattice geometry (as in some studies of tight-binding models~\cite{Voigt2001, Melin2005, Cheng2018, Zhou2024}) -- instead, it is encoded in the diagrammatic structure of the wavefunction.
The connectivity of the ZX-diagram imposes atypical subvolume-law scaling, which can be seen through the min-cut upper bound. Using MPS simulations, we find that the physical bipartite entanglement entropy scaling can be tuned by changing the free parameters of the ZX-diagrams: we give examples of the area law and logarithmic scaling for the triangle family, and scaling approximately as a logarithm and as a power law for the carpet family. In all cases, the entanglement profiles as well as local observables exhibit self-similar, fractal-like modulations. These clear non-thermal signatures make these fractal ZX-diagrams good candidates for QMBS.
For the triangle family, we construct a local parent Hamiltonian that is naturally represented in the same ZX language and for which the fractal ZX state is an exact eigenstate. This is done by combining standard methods with ZX-calculus insights, giving a set of local ZX identities that certify the annihilation of the target states by certain local Hamiltonian terms. After resolving all identified exact symmetries, we find near-Poissonian spectral statistics of this parent Hamiltonian. We further construct a perturbed local Hamiltonian, again compactly representable as a ZX-diagram, whose level statistics are clearly chaotic, while the same triangle-based fractal ZX state remains an exact eigenstate in the spectral bulk. Thus, the fractal ZX-diagram can be embedded as a quantum many-body scar in an otherwise chaotic Hamiltonian.

Our results suggest that ZX-calculus can be used as a diagrammatic route to finding exact quantum many-body scars and exact forms of the corresponding chaotic Hamiltonians, adding a new method to the existing landscape alongside algebraic-tower, projector-embedding, and MPS constructions.
The distinction from conventional tensor-network parent-Hamiltonian constructions is that the graphical language here does more than encode the target wavefunction. Local ZX identities certify the annihilation conditions directly and expose deformations that preserve the exact eigenstate while changing the spectral properties of the surrounding Hamiltonian.
ZX-calculus, therefore, acts here as an algebraic design language relating state preparation, local constraints, and Hamiltonian engineering, suggesting that it can serve more broadly as a construction language for quantum many-body physics.

This paper is structured as follows. We begin by introducing the necessary methods used in this paper in Sec.~\ref{sec:methods}. Next, we construct fractal ZX-diagrams in Sec.~\ref{sec:fractalZX} that represent many-body quantum states, and investigate their properties numerically. In Sec.~\ref{sec:hams} we construct the parent Hamiltonian as a ZX-diagram, and embed the fractal ZX state into the spectrum of the associated chaotic Hamiltonian. Finally, we conclude in Sec.~\ref{sec:conclusions}.

\section{Background and methods}
\label{sec:methods}

In this section, we briefly summarize the diagrammatic and numerical tools used throughout the paper.

\subsection{ZX-calculus}

The ZX-calculus is a graphical language for linear maps between qubits, which is now an established toolkit for quantum circuit simplification and quantum compilation~\cite{Coecke2011, Coecke2017, vandeWetering2020}. ZX-calculus is composed of ZX-diagrams, which represent the linear maps, and rewrite rules -- equivalence transformations between them. A ZX-diagram is an undirected graph, whose elementary generators are the Z-spiders (green vertices) and X-spiders (red vertices), together with ordinary wires. A Z-spider with $m$ incoming and $n$ outgoing legs and phase $\alpha$ denotes the tensor
\begin{equation}
    \begin{ZX}
        \leftManyDots[0.8][1]{m} \zxZ{\alpha} \rightManyDots[0.8][1]{n}
    \end{ZX}
    \,\, := \ket{0}^{\otimes n} \bra{0}^{\otimes m} + e^{i \alpha} \ket{1}^{\otimes n} \bra{1}^{\otimes m},
\end{equation}
and similarly for the X-spider,
\begin{equation}
    \begin{ZX}
        \leftManyDots[0.8][1]{m} \zxX{\alpha} \rightManyDots[0.8][1]{n}
    \end{ZX}
    \,\, := \ket{+}^{\otimes n} \bra{+}^{\otimes m} + e^{i \alpha} \ket{-}^{\otimes n} \bra{-}^{\otimes m}.
\end{equation}
When $\alpha=0$, the phase is omitted. Additionally, it is useful to define a Hadamard gate
\begin{equation}
    \begin{ZX}[column sep=1cm]
        \ar[r,H] &
    \end{ZX}
    =
    \begin{ZX}[column sep=1cm]
        \ar[r,blue,dashed] &
    \end{ZX}
    :=
    \begin{ZX}
        \ar[r] & \zxFracZ{\pi}{2} \rar & \zxFracX{\pi}{2} \rar & \zxFracZ{\pi}{2} \rar &
    \end{ZX},
\end{equation}
where the notation involving the blue dashed wire is often called the ``Hadamard edge''.
Equivalently, Z-spiders copy and match computational-basis information, while X-spiders do the same in the Hadamard-rotated basis. Also, Clifford gates and stabilizer states can be constructed using diagrams with ``Clifford phases'', where $\alpha$ is a multiple of $\pi/2$.
In this graphical notation, the composition of linear maps is represented by connecting wires, while tensor products are represented by juxtaposition. Thus, every closed or open ZX-diagram has an immediate interpretation as a tensor network.

A useful feature of the ZX notation is that the precise geometry of the drawing is not part of the data of the tensor network. Up to the ordering of boundary legs, only the connectivity, spider colors, and their phases matter; wires may be deformed without changing the underlying linear map. The diagrams may also be manipulated using local rewrite rules -- the complete rule set we summarize below~\cite{Vilmart2019},
\begin{subequations}
\begin{align}
    \begin{ZX}
        \leftManyDots{} \zxZ{\alpha} \middleManyDots{} & \zxZ{\beta} \rightManyDots{}
    \end{ZX}
    \overset{(\mathbf{f})}{=}
    \begin{ZX}
        \leftManyDots{} \zxZ{\alpha+\beta} \rightManyDots{}
    \end{ZX}&, \quad
    \begin{ZX}
        \leftManyDots{} \zxX{} \rar &
        \zxZ{} \rightManyDots{} &
    \end{ZX}\!
    \overset{(\mathbf{b})}{=}
    \begin{ZX}
        \rar & \zxZ{} \rar \ar[rd] \ar[d,3 vdots] & [\zxDotsCol] \zxX{} \rar \ar[d,3 vdots] & \\[\zxDotsRow]
        \rar & \zxZ{} \rar \ar[ru] & \zxX{} \rar & 
    \end{ZX},\\[1em]
    \begin{ZX}[column sep=0.25cm,row sep=0.25cm]%[zx row sep=5pt]
        \zxN{} \ar[rd,-N.,H={pos=.35}] &[\zxwCol,\zxHCol] &[\zxwCol,\zxHCol] \zxN{} \\[\zxNRow] \ar[r,3 vdots]%%
        & \zxX{\alpha} \ar[r,3 vdots]
        \ar[ru,N'-,H={pos=1-.35}]
        \ar[rd,N.-,H={pos=1-.35}] & \\[\zxNRow]
        \zxN{} \ar[ru,-N',H={pos=.35}] 
        & & \zxN{}
    \end{ZX}
    \overset{(\mathbf{c})}{=}
    \begin{ZX}[column sep=0.25cm,row sep=0.25cm]%[zx row sep=5pt]
        \zxN{} \ar[rd,-N.] &[\zxwCol,\zxHCol] &[\zxwCol,\zxHCol] \zxN{} \\[\zxNRow] \ar[r,3 vdots]%%
        & \zxZ{\alpha} \ar[r,3 vdots]
        \ar[ru,N'-]
        \ar[rd,N.-] & \\[\zxNRow]
        \zxN{} \ar[ru,-N'] 
        & & \zxN{}
    \end{ZX}, \ \  & \ \  
    \begin{ZX}[column sep=0.3cm]
        \rar & \zxZ{} \rar &
    \end{ZX}
    =
    \begin{ZX}[column sep=0.3cm]
        \rar & \zxX{} \rar &
    \end{ZX}
    \overset{(\mathbf{id})}{=}
    \begin{ZX}[column sep=0.3cm]
        \ar[rr] &&
    \end{ZX}, \\[1em]
    \begin{ZX}[column sep=0.2cm]
        \ar[r] & \zxZ*{\alpha_1} \rar & \zxX*{\alpha_2} \rar & \zxZ*{\alpha_3} \rar &
    \end{ZX}
    &\overset{(\mathbf{e})}{=}
    \begin{ZX}[column sep=0.2cm]
        \ar[r] & \zxX*{\beta_1} \rar & \zxZ*{\beta_2} \rar & \zxX*{\beta_3} \rar &
    \end{ZX}.
\end{align}\label{eq:ZXrules}
\end{subequations}
It includes: spider fusion (\textbf{f}), color change by Hadamard conjugation (\textbf{c}), bialgebra rule (\textbf{b}), identities (\textbf{id}), and the Euler rule (\textbf{e}) (phases in the Euler rule are Euler angles, with their precise definitions given in Ref.~\cite{Vilmart2019}). These rules are sound (if two diagrams are related by the rewrite rules, then they represent the same linear map) and complete (any two diagrams that represent the same linear map can be deformed into each other using the above rule set)~\cite{Ng2017, Jeandel2018, Backens2014, Backens2014CT}.

In the present work, we use ZX-diagrams in two complementary ways. First, they define the many-body wavefunctions themselves. A diagram with $L$ open boundary legs on the right is interpreted as an unnormalized state on $L$ qubits. Global scalar factors will be neglected, as they are irrelevant for the calculation of expectation values and entanglement entropy. Second, the diagrams provide a compact symbolic representation of local maps and Hamiltonian terms.
For practical manipulation of diagrams, we use ZXLive~\cite{ZXLive}, an interactive graphical interface for ZX-diagrams, whose backend is PyZX~\cite{Kissinger2020PyZX}. PyZX is a Python library that implements automated simplification and rewriting routines for large ZX-diagrams, including the standard stabilizer and graph-like simplification methods. We use these tools to check diagrammatic identities, simplify intermediate expressions, and convert between diagrammatic and circuit-like representations when useful.

\subsection{Matrix product states and numerical compression}
\label{sec:methods:mps}

To compute entanglement entropies and local observables for the fractal ZX states, we translate the corresponding many-body wavefunctions to the matrix product state (MPS) formalism~\cite{Schollwock2011, Verstraete2008, Vidal2004, Cirac2021}. An MPS for an $L$-qubit state has the form
\begin{equation}
    \ket{\psi} = \sum_{\sigma_1,\ldots,\sigma_N} A^{[1]\sigma_1} A^{[2]\sigma_2} \cdots A^{[L]\sigma_L} \ket{\sigma_1\cdots\sigma_L},
\end{equation}
where the bond dimensions of the matrices $A^{[i]\sigma_i}$ control the amount of entanglement retained across each bipartition, and $\ket{\sigma_1\cdots\sigma_L}$ are computational basis states. This representation is especially well-suited to one-dimensional cuts through the output legs of our diagrams: the von Neumann entropy for every contiguous bipartition can be obtained from the Schmidt values on the corresponding MPS bond.

The state-preparation procedure is as follows. We first rewrite each ZX-diagram as a sequence of elementary operations acting on an initial product state, $|+\rangle^{\otimes L} = \{\begin{ZX} \zxZ{} \ar[rr] && \end{ZX} \}^{\otimes L}$ (which can be obtained by unfusing phaseless Z-spiders on the left side of the diagram). Operationally, the diagram is decomposed into a circuit-like tensor network consisting of one- and two-qubit unitary gates (controlled-Z gates, Hadamard gates, Z-phase gates), as well as local projectors $\ket{+}\bra{+}$. Note that although efficient circuit extraction is, in general, a hard problem for circuits with no postselection~\cite{Backens2021}, here we will deal with ZX-diagrams that not only already have a circuit-like structure, but we will also allow postselected measurements (projectors). This nonunitary circuit is then applied sequentially to an MPS representing the all-plus $|+\rangle^{\otimes L}$ state. After each application of a gate or a projector, the MPS is compressed by singular-value decomposition and normalized~\cite{Vidal2004}. The compression is controlled by a maximum bond dimension $\chi_{\max}$ and a discarded-weight cutoff $\epsilon_{\rm SVD}$. Convergence is checked by increasing $\chi_{\max}$ and decreasing $\epsilon_{\rm SVD}$ until the reported entanglement entropies and local observables are stable on the scale of the plotted data.

All MPS calculations are performed using the Python tensor-network library quimb~\cite{Gray2018}, which provides routines for applying local gates and projectors, computing entanglement entropy, and evaluating expectation values of local observables, as well as calculating local reduced density matrices. Specifically, for a normalized MPS $\ket{\psi}$, the bipartite von Neumann entanglement entropy across the cut between sites $j$ and $j+1$ is computed from the Schmidt coefficients $\{\lambda_a^{(j)}\}$ as
\begin{equation}
    S = -\sum_a (\lambda_a^{(j)})^2 \log_2 (\lambda_a^{(j)})^2.
\end{equation}
The same MPS representation also allows efficient evaluation of one- and few-site expectation values,
\begin{equation}
    \langle O_i\rangle = \langle \psi|O_i|\psi\rangle ,
\end{equation}
as well as reduced density matrices
\begin{equation}
    \rho_A = {\rm Tr}_{\bar A} |\psi\rangle\langle\psi|
\end{equation}
for small subsystems $A$. $\bar A$ appearing in the partial trace is the complement of subsystem $A$. These reduced density matrices will be used for the parent-Hamiltonian construction described below.

\subsection{Parent Hamiltonian construction}
\label{sec:methods:parHam}

Here, we briefly describe the construction of a local parent Hamiltonian $H = \sum_i h_i$ for which a target state $|\psi\rangle$ is an exact ground state~\cite{Affleck1987, Fannes1992, PerezGarcia2007, Cirac2021}.
First, we choose the range $k$ for the local Hamiltonian terms. Next, we construct reduced density matrices for each of $k$ adjacent sites, $\rho_{R_i} = {\rm Tr}_{\bar{R_i}} |\psi\rangle\langle\psi|$, where $R_i$ is the region of $k$ adjacent sites starting from site $i$. Any positive semidefinite operator whose support lies inside the kernel of $\rho_{R_i}$ annihilates the target state locally. A canonical choice for the local Hamiltonian is the projector onto the kernel,
\begin{equation}
    h_i = \sum_{\{\lambda_a=0\}} \ket{v_a}\bra{v_a},
\end{equation}
where $\lambda_a$ and $\ket{v_a}$ are eigenvalues and eigenvectors of $\rho_{R_i}$. The parent Hamiltonian $H$ constructed this way is local, frustration-free (each $h_i$ shares the common ground state), and has $\ket{\psi}$ as a zero-energy ground state.
The interaction range $k$ is chosen large enough so that $\rho_{R_i}$ has a nontrivial kernel; in practice, we scan over local ranges and choose the smallest range for which null vectors appear.

For scar constructions, it is often useful to generalize this parent-Hamiltonian construction. Instead of requiring that the target state be a ground state of a positive frustration-free Hamiltonian, one may construct local annihilators and embed the state into the spectrum of a more general Hamiltonian. A common form is
\begin{equation}
    H = \sum_i h_i V_i h_i + H_{\rm aux},
\end{equation}
where $V_i$ are otherwise generic local Hermitian operators, and $H_{\rm aux}$ acts diagonally or uniformly on the target subspace. Such projector-embedding constructions provide a systematic way of embedding exact nonthermal eigenstates in otherwise generic spectra~\cite{Shiraishi2017, Moudgalya2018, Moudgalya2020}.
We will exploit the freedom of choosing $V_i$ to search for Hamiltonian terms that are not only local but also compactly representable as ZX-diagrams.

\subsection{Hamiltonian spectral statistics}
\label{sec:methods:spectrum}

This subsection describes the spectral diagnostics used to classify Hamiltonians as integrable or chaotic and the practical procedures used to make those diagnostics robust~\cite{Berry1977, Bohigas1984, DAlessio2016}.

For an ordered many-body spectrum $E_n$ we compute the ratio of consecutive level gaps~\cite{Oganesyan2007}
\begin{equation}
    r_n=\min(\tilde r_n,\tilde r_n^{-1}), \quad \tilde r =\frac{E_{n+1}-E_n}{E_n-E_{n-1}}.
\end{equation}
$r$ is insensitive to local density variations and does not require spectral unfolding. We will specifically be interested in the probability density $P(r)$ and the average value $\langle r \rangle$. To account for finite-size effects near the edges of the energy spectrum, we will only use the middle 80\% of energy levels. Before performing the statistics, one also needs to resolve all symmetries of the Hamiltonian.

For integrable models, energy levels are uncorrelated, and the level statistics is Poissonian, $P_{\text{P}}(r) = 2 / (1 + r)^2$, and its average value is $\langle r \rangle_{\text{P}} = 2\ln 2-1 \approx 0.38629$. For chaotic Hamiltonians, there is a dip of $P(r)$ near $r=0$ indicating energy level repulsion, and the spectrum is described by Wigner-Dyson statistics, for which an approximate expression is~\cite{Atas2013}
\begin{align}
    P(r) &\approx \frac{2}{Z_{\beta}} \frac{(r+r^2)^{\beta}}{(1+r+r^2)^{1+3\beta/2}} \\
    &\quad+ \frac{2 C_{\beta}}{(1+r)^2} \left( \Big(r+\frac{1}{r}\Big)^{-\beta} - c_{\beta} \Big(r+\frac{1}{r}\Big)^{-\beta-1} \right),\nonumber
\end{align}
where $\beta$ is the Dyson index controlling the power of the eigenvalue repulsion. If an antiunitary symmetry squaring to $+1$ acts within the resolved symmetry sector of the Hamiltonian (usually this is the time-reversal symmetry), the appropriate ensemble is the Gaussian Orthogonal Ensemble (GOE, $\beta=1$); if no antiunitary symmetry acts internally within that sector, the Gaussian Unitary Ensemble (GUE, $\beta=2$) should be used. Specifically, for GOE we have parameters $Z_{\beta=1} = 8/27$, $C_{\beta=1} \approx 0.233378$, and $c_{\beta=1}=2(\pi-2)/(4-\pi)$, while for GUE $Z_{\beta=2} = (4\pi)/(81\sqrt{3})$, $C_{\beta=2} \approx 0.578846$, and $c_{\beta=2}=4(4-\pi)/(3\pi-8)$. The average values are $\langle r \rangle_{\text{GOE}} \approx 0.5307(1)$ and $\langle r \rangle_{\text{GUE}} \approx 0.5996(1)$.

Interestingly, because the level statistics probes eigenvalue correlations rather than eigenstate structure, they often detect the emergence of chaos even in relatively small systems, making $\langle r \rangle$ one of the most sensitive diagnostics of integrability breaking available in exact diagonalization studies. Deviations from either limiting value can reveal finite-size effects, approximate symmetries, or the presence of intermediate dynamical regimes.

\section{Fractal ZX-diagrams}
\label{sec:fractalZX}

In this section, we construct two families of pure many-body states by interpreting the graphs of the Sierpi\'nski triangle and Sierpi\'nski carpet as ZX-diagrams; the resulting states inherit the self-similar structure of the underlying fractals. The reason why such a construction would be particularly interesting lies in the graph-theoretic upper bounds on their bipartite entanglement, which use minimum cuts of the underlying fractal connectivity -- both families exhibit entanglement that grows parametrically more slowly than the volume law expected for generic finite-energy-density eigenstates. Furthermore, we calculate the actual entanglement entropy and local observables using matrix-product-state simulations; these properties retain self-similar spatial structures when increasing the system size, and make the fractal ZX states natural candidates for atypical eigenstates of interacting many-body Hamiltonians. Interestingly, we will be able to tune the specific entanglement scaling by changing the free parameters of the ZX-diagram. Low entanglement and structured local observables alone, however, are not sufficient to identify a quantum many-body scar: the states must also be embedded at finite energy density in an otherwise thermal spectrum. The Hamiltonian construction for these states will be addressed in Sec.~\ref{sec:hams}.

\subsection{Fractal ZX-diagram construction}

\begin{figure}[t]
    \centering
    \includegraphics[width=\columnwidth]{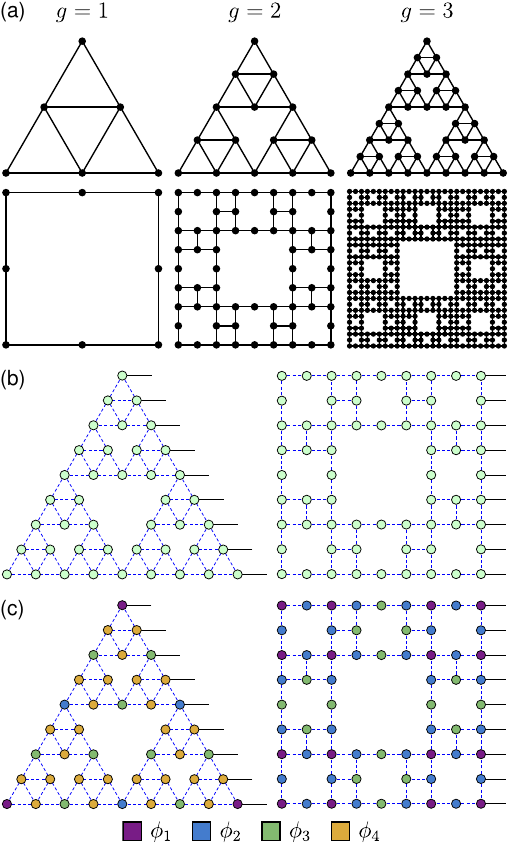}
    \caption{(a)~Successive generations $g$ for graphs of the Sierpi\'nski triangle (top row) and the Sierpi\'nski carpet (bottom row). (b)~Example fractal ZX-diagrams constructed from the triangle $g=3$ and the carpet $g=2$. Spider phases are omitted for clarity. (c)~Generation-dependent spider phases $\phi_g$ in the ZX-diagrams are indicated here by color.}
    \label{fig:generations}
\end{figure}

The construction begins with a Sierpi\'nski-triangle or Sierpi\'nski-carpet graph of a chosen generation $g$, where $g$ counts the number of recursive steps used to generate the fractal geometry; examples are shown in Fig.~\ref{fig:generations}(a). We promote this graph to a ZX-diagram by placing a Z-spider at every vertex, replacing every graph edge by a Hadamard edge, and assigning generation-dependent phases to the Z-spiders. Such ZX-diagrams, where the only spiders are Z-spiders, and the only non-boundary edges are Hadamard edges, are often called graph-like ZX-diagrams~\cite{Duncan2020}. Finally, boundary legs are attached to the vertices along the right side of the fractal, converting the graph into a many-body state, as illustrated in Fig.~\ref{fig:generations}(b).

For the Sierpi\'nski triangle, all spiders introduced at generation $g$ are assigned the same phase $\phi_{g+1}$. For the Sierpi\'nski carpet, spiders introduced at generation $g$ are assigned either $\phi_g$ or $\phi_{g+1}$, depending on whether the corresponding vertex lies at a corner of a square or at the midpoint of one of its sides. The resulting phases are illustrated in Fig.~\ref{fig:generations}(c). In the numerical calculations, we consider four representative phase families: $\phi_g = \pi/g$, $\phi_g = \pi/2^{g-1}$, $\phi_g = \pi(1-1/g)$, and $\phi_g = \pi(1-1/2^{g-1})$. These four choices control (i)~the rate at which phases change with generation (inverse-linear vs exponential) and (ii)~whether phases decrease toward 0 or increase toward $\pi$ as $g$ grows. In all cases, these phases approach Clifford angles (multiples of $\pi/2$) despite consisting mostly of non-Clifford phases. To probe the phase dependence beyond these families, we also consider non-Clifford and oscillatory phase functions, including $\phi_g=\pi/3$ and $\phi_g=\pi/3+\pi/2^{2g-2}$ for the triangle and $\phi_g=\pi(1+(-1)^g)/8$ for the carpet.

The resulting diagrams represent pure states of $L$ qubits, with $L=2^g+1$ for the triangle and $L=3^g$ for the carpet. Because both the graph and the phase assignment are reflection symmetric, the corresponding many-body states possess an exact mirror symmetry.

\subsection{Entanglement bound from the minimum cut}

A first indication of the entanglement structure follows directly from the connectivity of the ZX-diagrams. Specifically, we use the minimum-cut entanglement entropy, which is a graph-theoretic upper bound on the entanglement. For a bipartition of the external legs (which correspond to qubits), one finds the minimal number of internal wires that must be cut to separate the two sets of boundary legs. Since each cut wire carries a bond dimension of two, each can at most increase the entanglement entropy by 1. This minimum-cut entropy depends only on the connectivity of the tensor network -- it is consequently independent of the spider phases and does not account for algebraic cancellations, destructive interference, or ZX-calculus simplifications. It should therefore be interpreted as the \textit{entanglement capacity} of the diagram rather than as a prediction of the entropy of the represented state.

\begin{figure}[t]
    \centering
    \includegraphics[width=\columnwidth]{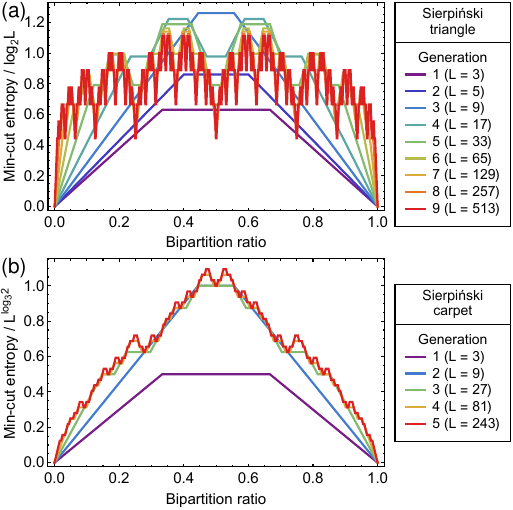}
    \caption{Minimum-cut entanglement bound for states represented by ZX-diagrams based on (a)~the Sierpi\'nski triangle and (b)~the Sierpi\'nski carpet, for successive generations $g$ (and the corresponding system sizes $L$). The maximum min-cut grows logarithmically with system size $L$ for the triangle and as a power law $\sim L^{\log_3 2}$ for the carpet.}
    \label{fig:mincut}
\end{figure}

Fig.~\ref{fig:mincut}(a) shows the minimum-cut entropy for the Sierpi\'nski triangle, rescaled by $\log_2 L$. The maximum of min-cut entropy increases linearly with the generation $g$ (more precisely, it is equal to $g + 1$ for $g \ge 3$) and therefore asymptotically it increases as a logarithm of the number of qubits $L$. We can understand this scaling geometrically: the minimum cut successively passes through the gaps created at different levels of the fractal hierarchy. Because the linear dimensions of these gaps increase geometrically with the generation, only a number of wires proportional to the number of hierarchical levels must be cut. This logarithmic bound is reminiscent of the entanglement scaling encountered in one-dimensional critical ground states. Here, however, it is only an upper bound derived from the diagrammatic connectivity and should not by itself be interpreted as evidence of criticality.

The Sierpi\'nski carpet ZX-diagram displays a qualitatively different behavior. As shown in Fig.~\ref{fig:mincut}(b), the minimum-cut curve (as a function of the bipartition) collapses to a fractal-like pyramid shape for large systems when rescaled by $L^{\log_3 2}$. This suggests a power-law entanglement capacity scaling, with a specific exponent of $\log_3 2 \approx 0.6309$. Again, we can understand this through the geometry of the underlying graph: the discrete Sierpi\'nski carpet at iteration $g$ decomposes recursively into 8 smaller carpets, of which two need to be traversed by the min-cut starting near the middle of the carpet side. Hence, the minimum number of edges you must cut doubles at every iteration, leading to the min-cut entropy growth of $2^g = L^{\log_3 2}$. This also matches the Hausdorff dimension of the cross-section of the carpet (a Cantor set)~\cite{Zhou2024}. Intuitively, the min-cut must cross many smaller-scale structures before it can take advantage of a square-shaped gap at a larger scale; thus, the denser carpet connectivity leads to faster entanglement scaling than that of the triangle. Power-law/fractal entanglement scaling is uncommon in practice -- cases that do appear are generally linked to anomalous transport (e.g., KPZ-type dynamics), fractal operator support or geometry, or to fine-tuned constraints or symmetries that restrict local dynamics~\cite{Bulchandani2021, Zhou2024}.

Importantly, these results show that a volume-law entropy is impossible for any contiguous bipartition of these fractal ZX-diagrams. The states are therefore kinematically restricted to a nonthermal entanglement scaling, independently of the particular choice of phases. We next determine how the entanglement scaling depends on the spider phase function within these graph-theoretic bounds.

\subsection{Entanglement entropy}

\begin{figure}[t]
    \centering
    \includegraphics[width=\columnwidth]{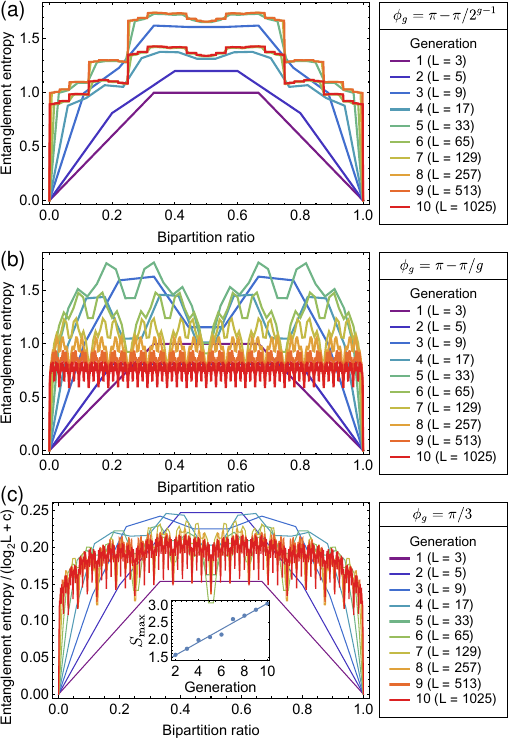}
    \caption{The von Neumann entanglement entropy as a function of the bipartition ratio for successive generations $g$ (and the corresponding system sizes $L$) for the Sierpi\'nski triangle ZX-diagram. Z-spider phases are (a)~$\phi_g = \pi(1-1/2^{g-1})$, (b)~$\phi_g = \pi(1-1/g)$, and (c)~$\phi_g = \pi/3$ (constant used to rescale the entropy is $c=4$). The inset in (c) shows the dependence of the maximum entanglement across all bipartitions $S_\text{max}$ vs $g$ for $\phi_g = \pi/3$.}
    \label{fig:entanglement}
\end{figure}

To calculate the actual entanglement entropy and local observables, we convert the fractal ZX-diagrams into matrix product states, using the procedure described in Sec.~\ref{sec:methods:mps}. Specifically, the ZX-diagrams in question already admit a circuit-like form on $L$ qubits in Fig.~\ref{fig:generations}(b). For the MPS compression, we find that the results are sufficiently converged for the maximum bond dimension of $\chi_{\max} = 2048$ and the cutoff of $\epsilon_{\rm SVD} = 10^{-20}$.

Fig.~\ref{fig:entanglement} shows the von Neumann entropy of contiguous bipartitions of the Sierpi\'nski-triangle ZX-diagram for a few functional forms of $\phi_g$. For the $\phi_g = \pi(1-1/2^{g-1})$ family of diagrams [panel (a)], the entropy develops a self-similar spatial profile: at increasing system size, the curves approach two distinct limiting fractal patterns for odd and even $g$. Panel (b) shows the $\phi_g = \pi(1-1/g)$ family of diagrams, where although we observe very clear signatures of fractality, the entropy seems to decrease (to presumably zero in the thermodynamic limit) as the number of qubits increases. The other two families of near-Clifford phase spiders show similar signatures (results have been relegated to Appendix~\ref{sec:appendix-data}), with $\phi_g = \pi/2^{g-1}$ showing convergence to entropy of $S=1$, while $\phi_g = \pi / g$ produces a similar behavior to that of $\phi_g = \pi(1-1/g)$. Across all four near-Clifford phase families, we observe that the bipartite entanglement entropy of contiguous cuts remains bounded by an $O(1)$ constant as $L$ increases.
These states, therefore, obey an area law, substantially below the logarithmic upper bound inferred from the min-cut picture.

However, when using a constant phase family of $\phi_g = \pi/3$ [see panel (c)], we find entropy profiles that converge if rescaled by $\log L+c$ (with $c\approx 4$). Similar behavior can be seen in a non-constant phase family $\phi_g = \pi/3+\pi/2^{2g-2}$ (see Appendix~\ref{sec:appendix-data}). This implies that for certain phase families, entropy increases logarithmically with the system size, realizing the asymptotic scaling of the minimum-cut bound, and that changing the form of $\phi_g$ allows tuning the entanglement entropy scaling from area-law to logarithmic.

The Sierpi\'nski-carpet ZX-diagrams exhibit a similar sensitivity of the entanglement scaling to the choice of spider phases. For the four near-Clifford phase families, we observe much slower growth than the $L^{\log_3 2}$ scaling permitted by the minimum cut, consistent with an approximately logarithmic dependence over the accessible system sizes (although it may instead be a weak fractal scaling $S\sim L^{\log_3 2/2}$). A representative example is shown in Fig.~\ref{fig:entanglement_carpet}(a), with the remaining phase families presented in Appendix~\ref{sec:appendix-data}. In contrast, for the alternating phase family $\phi_g = \pi(1 + (-1)^g)/8$, shown in Fig.~\ref{fig:entanglement_carpet}(b), the entanglement grows substantially faster and is consistent with a power law $S\sim L^\alpha$, with $\alpha\approx1/2$. Although this remains below the minimum-cut exponent $\log_3 2 \approx 0.6309$, it demonstrates that the entanglement scaling of the carpet states is not fixed by their graph connectivity and can instead be controlled through the spider phase function, while also raising the possibility that a suitable phase choice may realize the full $L^{\log_3 2}$ minimum-cut scaling.

Taken together, these results reveal the spider phase function as a control parameter for the entanglement structure of the fractal ZX-diagrams. For the Sierpi\'nski triangle, we can obtain area-law and logarithmic entanglement scalings, while for the Sierpi\'nski carpet, changing the phase function similarly tunes the entropy from approximately logarithmic to a stronger power law. This provides a particularly simple mechanism for engineering scar candidates with different entanglement structures within a common diagrammatic family. Tunable entanglement has also recently emerged in other constructions of quantum many-body scars, including thermofield-double states with effective temperature as a control parameter~\cite{Teretenkov2024}, and symmetric-tensor scars with proportion of triplet flavors controlling the entanglement~\cite{Mukherjee2026}. Our construction realizes a complementary form of this control directly through free parameters of the diagrammatic ZX representation.

\begin{figure}[t]
    \centering
    \includegraphics[width=\columnwidth]{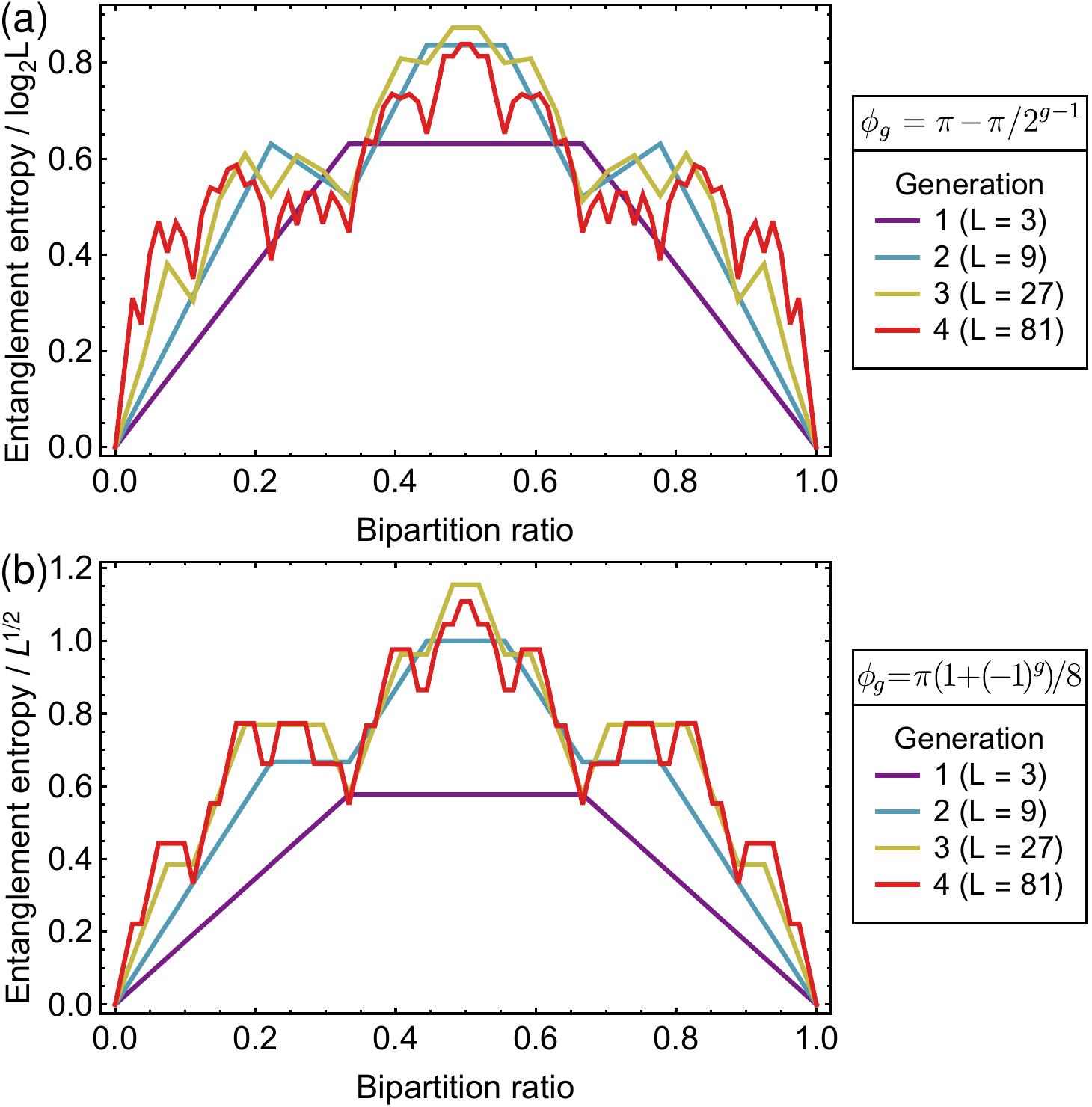}
    \caption{Entanglement entropy as a function of the bipartition ratio for different generations (and the corresponding system sizes) of the Sierpi\'nski carpet ZX-diagram. Z-spider phases are (a)~$\phi_g = \pi(1-1/2^{g-1})$, and (b)~$\phi_g = \pi(1+(-1)^g)/8$.}
    \label{fig:entanglement_carpet}
\end{figure}

\subsection{Local observables}

\begin{figure}[t]
    \centering
    \includegraphics[width=\columnwidth]{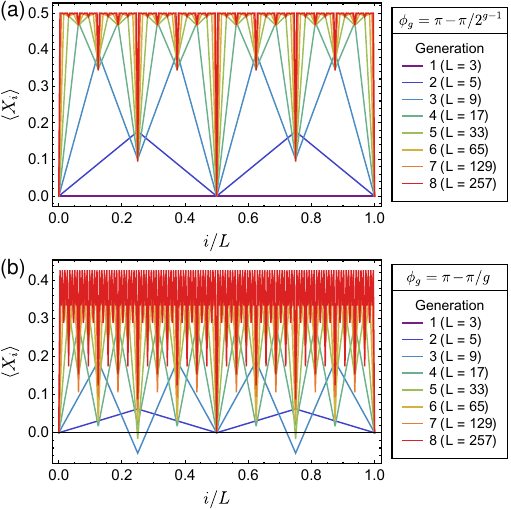}
    \caption{Local expectation value $\langle X_i \rangle$ as a function of the rescaled position $i/L$, for the Sierpi\'nski triangle ZX-diagram. Z-spider phases are (a)~$\phi_g = \pi(1-1/2^{g-1})$ and (b)~$\phi_g = \pi(1-1/g)$.}
    \label{fig:Xexpectation}
\end{figure}

Since entanglement provides a nonlocal diagnostic of atypicality, we complement it by examining local expectation values, which can more directly identify the nonthermal states. Fig.~\ref{fig:Xexpectation} shows the expectation value $\langle X_i\rangle$ for the Sierpi\'nski-triangle construction. For all four near-Clifford spider phase families (two are shown in Appendix~\ref{sec:appendix-data}), $\langle X_i\rangle$ displays abrupt fractal-like variations across the chain. These oscillations persist as the system size increases and reproduce nontrivial fractal structure on progressively finer spatial scales. By contrast, we find $\langle Z_i\rangle=0$ at every site, which agrees with the thermodynamic average. The persistent self-similar structure of $\langle X_i \rangle$ provides a local diagnostic of the atypical character of the state.

In summary, fractal ZX-diagrams constructed in this section possess subextensive entanglement entropy, controlled through ZX spider phase functions, and self-similar profiles of local expectation values. These features suggest that these are promising quantum many-body scar candidates. In the next section, we focus on the triangle ZX-diagram and find a canonical choice for both the parent and the chaotic Hamiltonians that admit it as a ground state and a quantum many-body scar, respectively.

\section{ZX construction of parent and chaotic Hamiltonians}
\label{sec:hams}

We now turn from the structure of the fractal ZX-diagram states to the Hamiltonians that support them. Given a tensor-network state, a standard route is to construct a local, frustration-free parent Hamiltonian from the kernels of its reduced density matrices~\cite{PerezGarcia2007}. Such a construction guarantees that the target state is a zero-energy ground state, although it does not by itself establish nonintegrability. A similar mechanism is used during the projector-embedding constructions for chaotic Hamiltonians with quantum many-body scars; there, local annihilators act nontrivially on the rest of the Hilbert space~\cite{Shiraishi2017}. This freedom can place an exact, atypical state inside an otherwise chaotic spectrum.

Here, these two related constructions acquire a particularly transparent graphical representation. We first use the MPS form of the fractal triangle ZX state to determine the minimal support of its local annihilators as well as their numerical form. (Preliminary attempts at the carpet ZX-diagram indicate that it requires local annihilators with considerably larger support, making this state less suitable for Hamiltonian engineering demonstration.) We then construct new annihilators using insights from ZX-calculus. ZX-calculus here specifically shows the local mechanism by which the fractal ZX state is annihilated. The resulting parent Hamiltonian can be expressed through a small set of simple ZX-diagrams. Moreover, the deformation that converts the near-integrable parent Hamiltonian into a chaotic one is itself represented by a modification of the local ZX-diagram representing a Hamiltonian term. Thus, the exact eigenstate, its parent Hamiltonian, and the chaotic Hamiltonian that hosts the state as a QMBS, all remain within the same graphical language.

\subsection{Frustration-free parent Hamiltonian}

We begin by constructing the parent Hamiltonian numerically from the kernels of reduced density matrices, following the procedure described in Sec.~\ref{sec:methods:parHam}. All required reduced density matrices and null spaces can be obtained efficiently from the MPS representation of the fractal ZX state. Specifically, we use the Sierpi\'nski triangle ZX-diagram with the spider phase family of $\phi_g = \pi(1-1/g)$.

The smallest intervals for which nontrivial local annihilators exist have ranges 3 and 5. Numbering the physical sites from $1$, the five-site terms are supported on intervals $\{3,4,5,6,7\} + 4m, m\in\mathbb{Z}$, whereas the three-site terms act on $\{1,2,3\} + 4m$, $\{2,3,4\} + 4m$, and $\{3,4,5\} + 4m$. The resulting interaction range therefore follows the recursive structure of the boundary of the Sierpi\'nski triangle. We additionally find that overlapping parent-Hamiltonian terms do not commute. The model is consequently not a commuting-projector Hamiltonian, even though it remains frustration-free. Noncommutativity alone, however, does not imply quantum chaos, and the spectral properties must be examined independently.

Next, we aim to use the insights from ZX-calculus, as well as the freedom of choosing $h'_i = h_i V_i h_i$ as an alternative parent Hamiltonian, to construct a parent Hamiltonian with clear ZX structure. 
Rather than retaining the numerical kernel projectors directly, we search within each local annihilator subspace for analytically simple operators that admit compact ZX representations. The numerical null spaces determine the required support and constrain the operator form, while phase cancellation in the ZX-diagram suggests suitable analytic representatives.
Consider the phase hierarchy adjacent to the open legs of the Sierpi\'nski-triangle diagram in Fig.~\ref{fig:generations}(b). A local annihilator centered on a spider with phase $\phi_k$ must first compensate this phase. Diagrammatically, this is achieved by adjoining a pair of spiders with phases $-\phi_k$ and $\phi_k$, i.e., $\begin{ZX}\ar[rr] && \zxZ*{-\phi_k} && \zxZ*{\phi_k} \ar[rr] &&\end{ZX}$. Spider fusion then removes the phase from the central spider and exposes a phaseless Z-spider. Next, we will assume that other sites can be annihilated by phase-independent annihilators composed of Pauli operators. This indeed turns out to be enough to find the desired forms of all the local Hamiltonian terms $h'_i$.

\allowdisplaybreaks
To write down the Hamiltonian terms, let us first define the following operators and their corresponding ZX-diagrams,
\begin{subequations}
\begin{align}
    \Pi^Z_{i,j} &= \frac{I - Z_{i} Z_{j}}{2} =
    \begin{ZX}[column sep={0.5cm,between origins},row sep={0.5cm,between origins}]
        \ar[r] & \zxZ{} \ar[r] \ar[d] & \\
               & \zxX{\pi} \ar[d]& \\
        \ar[r] & \zxZ{} \ar[r] &
    \end{ZX}, \\[1em]
    \Pi^{XYZ}_{i,j} &= \frac{I - X_{i} X_{j} - Y_{i} Y_{j} - Z_{i} Z_{j}}{2} =
    \begin{ZX}[column sep={0.5cm,between origins},row sep={0.5cm,between origins}]
        \ar[r] & \zxZ{\pi}\ar[d] & \zxZ{\pi} \ar[r] \ar[d] & \\
        \ar[r] & \zxX{\pi}       & \zxX{\pi} \ar[r] &
    \end{ZX}, \\[1em]
    \Xi_i(k) &= \begin{pmatrix}
        1 & \exp{(-i \phi_k)} \\ \exp{(i \phi_k)} & 1
    \end{pmatrix} =
    \begin{ZX}[column sep={0.25cm},row sep={0.5cm,between origins}]
    \ar[r] & \zxZ*{-\phi_k} & \zxZ*{\phi_k} \ar[r] &\end{ZX}.
\end{align}\label{eq:projectors}%
\end{subequations}
Here, $\Pi^Z_{i,j}$ is the projector onto the odd-$Z$-parity subspace, the operator $\Pi^{XYZ}_{i,j}$ is twice the singlet projector, while $\Xi_i(k)$ is twice the rank-one projector onto $(\ket{0}+e^{i \phi_k}\ket{1})/\sqrt{2}$. Thus, all three operators are positive semidefinite, although the latter two are not normalized to be idempotent.

The local parent-Hamiltonian terms can then be constructed as,
\begin{subequations}
\begin{align}
    H^a_i(k) &= \Pi^Z_{i-1,i+1} \, \Xi_i(k) =
    \begin{ZX}[column sep={0.6cm,between origins},row sep={0.5cm,between origins}]
        \ar[rr] & & \zxZ{} \ar[rr] \ar[d] && \\
        \ar[r]  & \zxZ*{-\phi_k} & \zxX{\pi} \ar[d]& \zxZ*{\phi_k} \ar[r]& \\
        \ar[rr] & & \zxZ{} \ar[rr] &&
    \end{ZX},\label{eq:parentHam:Ha}\\[1em]
    H^b_i(k) &= \Pi^{XYZ}_{i-1,i+1}\, \Xi_i(k) =
    \begin{ZX}[column sep={0.5cm,between origins},row sep={0.5cm,between origins}]
        \ar[rr] & & \zxZ{\pi}\ar[dd] & \zxZ{\pi}\ar[rr] \ar[dd] && \\
        \ar[r]  & \zxZ*{-\phi_k} & & & \zxZ*{\phi_k} \ar[r]& \\
        \ar[rr] & & \zxX{\pi} & \zxX{\pi} \ar[rr] &&
    \end{ZX},\\[1em]
    H^c_i(k) &= \Pi^{XYZ}_{i-1,i+1} \Pi^{Z}_{i-2,i+2} \, \Xi_i(k)\\
    &= \begin{ZX}[column sep={0.5cm,between origins},row sep={0.5cm,between origins}]
        \ar[rrr] &&& \zxZ{} \ar[dd] \ar[rrr] &&& \\
        \ar[rr] & & \zxZ{\pi}\ar[dd] & & \zxZ{\pi}\ar[rr] \ar[dd] && \\
        \ar[r]  & \zxZ*{-\phi_k} & & \zxX{\pi}\ar[dd] & & \zxZ*{\phi_k} \ar[r]& \\
        \ar[rr] & & \zxX{\pi} & & \zxX{\pi} \ar[rr] && \\
        \ar[rrr] &&& \zxZ{} \ar[rrr] &&&
    \end{ZX}. \nonumber
\end{align}\label{eq:parentHam}%
\end{subequations}
The operators appearing in each term act on disjoint sites and therefore commute. Because every operator is positive semidefinite, $H_i^a$, $H_i^b$, and $H_i^c$ are themselves positive semidefinite. Once their annihilation of the fractal state is established, the sum of these terms is therefore manifestly frustration-free.

Importantly, the annihilation of the target state can be verified \textit{locally} in the ZX-diagram, without contracting the complete many-body wave function. One inserts the appropriate Hamiltonian term diagram into the corresponding boundary subdiagram of the state and applies the ZX rewrite rules (see Appendix~\ref{sec:appendix-annihilation} for an example). For the three types of local Hamiltonian terms, the relevant identities are
\begin{equation}
    \begin{ZX}[column sep={0.3cm,between origins},row sep={0.6cm,between origins}]
        &&&\ar[dr]&&&&&&&&&&&&\\
        &&\ar[rr]&&\zxZ*{\alpha} \ar[dr,blue,dashed] \ar[dl,blue,dashed] \ar[rrrrrrr] &&&&&&& \zxZ{} \ar[rrrr] \ar[d] &&&&\\
        &&& \zxZ*{\phi_k} \ar[dl,blue,dashed] \ar[dr,blue,dashed] \ar[rr,blue,dashed] && \zxZ*{\phi_k} \ar[dl,blue,dashed] \ar[dr,blue,dashed] \ar[rrrr] &&&&\zxZ*{-\phi_k}&&\zxX{\pi}\ar[d]&&\zxZ*{\phi_k} \ar[rr]&&\\
        \ar[rr]&&\zxZ*{\beta}\ar[rr,blue,dashed] \ar[dl]&&\zxZ*{\phi_k}\ar[rr,blue,dashed]&&\zxZ*{\gamma} \ar[rrrrr] \ar[dl] \ar[dr]&&&&&\zxZ{}\ar[rrrr]&&&&\\
        &&&&&&&&&&&&&&&
    \end{ZX} = 0,
    \label{eq:annihilHa}
\end{equation}
\begin{equation}
    \begin{ZX}[column sep={0.3cm,between origins},row sep={0.6cm,between origins}]
        &&&&&\ar[dr]&&&&&&&&&&&&&&&& \\
        &&&&\ar[rr]&&\zxZ*{\alpha} \ar[rr] \ar[dr,blue,dashed] \ar[dl,blue,dashed] &&&&&&&&&&&&&&& \\
        &&&&&\zxZ*{\beta} \ar[dr,blue,dashed] \ar[dl,blue,dashed] \ar[rr,blue,dashed]&&\zxZ*{\beta} \ar[dr,blue,dashed] \ar[dl,blue,dashed] \ar[rrrrrrrr]&&&&&&&&\zxZ*{\pi}\ar[dd]&&\zxZ*{\pi}\ar[dd] \ar[rrrr]&&&& \\
        &&&&\zxZ*{\phi_k} \ar[dr,blue,dashed] \ar[dl,blue,dashed] \ar[rr,blue,dashed]&&\zxZ*{\beta} \ar[rr,blue,dashed]&&\zxZ*{\phi_k} \ar[dr,blue,dashed] \ar[dl,blue,dashed] \ar[rrrrr]&&&&&\zxZ*{-\phi_k}&&&&&&\zxZ*{\phi_k}\ar[rr]&& \\
        &&&\zxZ*{\beta} \ar[dr,blue,dashed] \ar[dl,blue,dashed] \ar[rr,blue,dashed]&&\zxZ*{\beta} \ar[dr,blue,dashed] \ar[dl,blue,dashed]&&\zxZ*{\beta} \ar[dr,blue,dashed] \ar[dl,blue,dashed] \ar[rr,blue,dashed]&&\zxZ*{\beta} \ar[dr,blue,dashed] \ar[dl,blue,dashed] \ar[rrrrrr]&&&&&&\zxX*{\pi}&&\zxX*{\pi}\ar[rrrr]&&&& \\
        \ar[rr]&&\zxZ*{\gamma} \ar[dl] \ar[rr,blue,dashed]&&\zxZ*{\beta} \ar[rr,blue,dashed]&&\zxZ*{\phi_k} \ar[rr,blue,dashed]&&\zxZ*{\beta} \ar[rr,blue,dashed]&&\zxZ*{\delta} \ar[rr] \ar[dr] \ar[dl]&&&&&&&&&&& \\
        &&&&&&&&&&&&&&&&&&&&&
    \end{ZX} = 0,
\end{equation}
\begin{equation}
    \begin{ZX}[column sep={0.3cm,between origins},row sep={0.6cm,between origins}]
        &&&\ar[dr]&&&&&&&&&&&&&&&&& \\
        &&\ar[rr]&&\zxZ*{\alpha} \ar[dl,blue,dashed] \ar[dr,blue,dashed] \ar[rrrrrrrrrr]&&&&&&&&&&\zxZ*{}\ar[dd]\ar[rrrrrr]&&&&&& \\
        &&&\zxZ*{\beta} \ar[dl,blue,dashed] \ar[dr,blue,dashed] \ar[rr,blue,dashed]&&\zxZ*{\beta} \ar[dl,blue,dashed] \ar[dr,blue,dashed] \ar[rrrrrrr]&&&&&&&\zxZ*{\pi}&&&&\zxZ*{\pi}\ar[rrrr]&&&& \\
        \ar[rr]&&\zxZ*{\alpha} \ar[rr,blue,dashed]&&\zxZ*{\beta} \ar[rr,blue,dashed]&&\zxZ*{\phi_k} \ar[rrrr]&&&&\zxZ*{-\phi_k}&&&&\zxX*{\pi}&&&&\zxZ*{\phi_k}\ar[rr]&& \\
        &\ar[ur]&&&&\zxZ*{\beta} \ar[ur,blue,dashed] \ar[rr,blue,dashed]&&\zxZ*{\beta} \ar[ul,blue,dashed] \ar[rrrrr]&&&&&\zxX*{\pi}\ar[uu]&&&&\zxX*{\pi} \ar[rrrr] \ar[uu] &&&& \\
        &&\ar[rr]&&\zxZ*{\alpha} \ar[ur,blue,dashed] \ar[rr,blue,dashed]&&\zxZ*{\beta} \ar[ur,blue,dashed] \ar[ul,blue,dashed] \ar[rr,blue,dashed]&&\zxZ*{\alpha} \ar[ul,blue,dashed] \ar[rrrrrr]&&&&&&\zxZ*{}\ar[rrrrrr] \ar[uu]&&&&&& \\
        &&&\ar[ur]&&&&\ar[ur]&&\ar[ul]&&&&&&&&&&&
    \end{ZX} = 0.
\end{equation}
These identities establish, for the appropriate choices of the central site $i$, $H_i^a(k) \ket{\psi} = 0$, $H_i^b(k) \ket{\psi} = 0$, and $H_i^c(k) \ket{\psi} = 0$.

This diagrammatic proof also reveals why the construction is considerably more general than the numerical calculation from which it was inferred. The numerics were performed for the particular sequence $\phi_g=\pi(1-1/g)$, whereas the ZX identities depend only on the local simplification of $\phi_k$ spiders. Consequently, the same Hamiltonian form applies to an \textit{arbitrary} generation-dependent phase function, provided the local Hamiltonian term is assigned the phase of the corresponding boundary spider in the fractal ZX-diagram. Thus, varying the spider phase function can change the entanglement scaling of the exact eigenstate without altering the local annihilation mechanism underlying its Hamiltonian embedding.

The parent Hamiltonian $H_{\text{par}}$ can thus be formed by summing the local terms in Eq.~\eqref{eq:parentHam} by arranging them according to the periodic pattern $\{a, b, a, c, a, b, a, c, ...\}$, while their phases correspond to the phases found on the side of the Sierpi\'nski triangle ZX-diagram. As an example, for the generation $g=4$ state on $L=17$ sites, the corresponding Hamiltonian is
\begin{align}
    H_{\text{par}}^{g=4} &= H^a_2(5) + H^b_3(4) + H^a_4(5) + H^c_5(3)\\
    &\quad + H^a_6(5) + H^b_7(4) + H^a_8(5) + H^c_9(2)\nonumber\\
    &\quad + H^a_{10}(5) + H^b_{11}(4) + H^a_{12}(5) + H^c_{13}(3)\nonumber\\
    &\quad + H^a_{14}(5) + H^b_{15}(4) + H^a_{16}(5).\nonumber
\end{align}
Since all local terms are positive semidefinite, so is the Hamiltonian, and the fractal ZX state is an exact zero-energy ground state of $H_{\text{par}}$.

There is considerable freedom in the choice of the parent Hamiltonian terms. In particular, $H_i^b(k)$ annihilates the fractal state not only at the sites at which it appears in $H_{\mathrm{par}}$, but also at the even sites carrying terms of type $H_i^a(k)$. This follows from the annihilation identity
\begin{equation}
    \begin{ZX}[column sep={0.3cm,between origins},row sep={0.6cm,between origins}]
        &&&\ar[dr]&&&&&&&&&&&&&&\\
        &&\ar[rr]&&\zxZ*{\alpha} \ar[dr,blue,dashed] \ar[dl,blue,dashed] \ar[rrrrrrr] &&&&&&& \zxZ{\pi} \ar[dd] && \zxZ{\pi} \ar[rrrr] \ar[dd] &&&&\\
        &&& \zxZ*{\phi_k} \ar[dl,blue,dashed] \ar[dr,blue,dashed] \ar[rr,blue,dashed] && \zxZ*{\phi_k} \ar[dl,blue,dashed] \ar[dr,blue,dashed] \ar[rrrr] &&&&\zxZ*{-\phi_k}&&&&&&\zxZ*{\phi_k} \ar[rr]&&\\
        \ar[rr]&&\zxZ*{\beta}\ar[rr,blue,dashed] \ar[dl]&&\zxZ*{\phi_k}\ar[rr,blue,dashed]&&\zxZ*{\gamma} \ar[rrrrr] \ar[dl] \ar[dr]&&&&&\zxX{\pi}&&\zxX{\pi}\ar[rrrr]&&&&\\
        &&&&&&&&&&&&&&&&&
    \end{ZX} = 0.
    \label{eq:annihilHb_on_Ha_sites}
\end{equation}
Similarly, each $H_i^c(k)$ term annihilates the state at the sites carrying the $H_i^b(k)$ term, since the ZX-diagram for $H_i^b(k)$ is a subdiagram of $H_i^c(k)$. Consequently, the parent Hamiltonian may be chosen from a broad family of local terms that share the same null space on the fractal state. We will exploit this freedom in the subsequent subsection to construct controlled perturbations. Here, our choice for $H_{\text{par}}$ is such that its local terms have minimal operator support and rich symmetry properties.

We now examine this symmetry structure. By construction, it inherits the reflection symmetry $R$ of the fractal ZX-diagram. Additionally, it also possesses edge symmetries $Z_1$ and $Z_L$, a U(1) symmetry generated on even sites by $Q_e = \sum_{i\text{ even}} U_i(g_i)$, as well as a discrete symmetry at odd sites, $Q_o = X_1 \, \prod_{i=3\text{, }i\text{ odd}}^{L-2} U_i(g_i)\, X_L$. Here, $g_i$ denotes the generation label of the boundary spider associated with qubit $i$, and $U_i(k) = \Xi_i(k)-I$ is a unitary operator. These operators obey the following set of nontrivial commutation relations: $[Q_o, Z_1] \neq 0$, $[Q_o, Z_L] \neq 0$, $[R, Z_1] \neq 0$, $[R, Z_L] \neq 0$, while all other commutators are zero. Reflection exchanges the two edge symmetries, $R Z_1 R = Z_L$, while $Q_o$ flips both edge quantum numbers, $\{Q_o, Z_1\} = \{Q_o, Z_L\} = 0$.
Although ordinary time-reversal symmetry is broken here, the parent Hamiltonian preserves an antiunitary symmetry $\mathcal A = \prod_i X_i K$, where $K$ denotes complex conjugation in the computational basis. $\mathcal A$ satisfies $\mathcal A^2=+1$ and leaves $R$ and $Q_e$ invariant but flips both edge quantum numbers.

Note the relatively simple form of Eq.~\eqref{eq:parentHam}. Generic reduced-density-matrix constructions produce local projectors that are rather represented as dense numerical matrices. Here, using ZX-calculus insights, we found three ZX-diagram patterns built from spider phase cancellation, parity projection, and singlet projection. The fractal ZX-diagram state, therefore, gives rise to a Hamiltonian whose local algebra is substantially simpler than one might infer from the full many-body wave function.

\begin{figure}[t]
    \centering
    \includegraphics[width=\columnwidth]{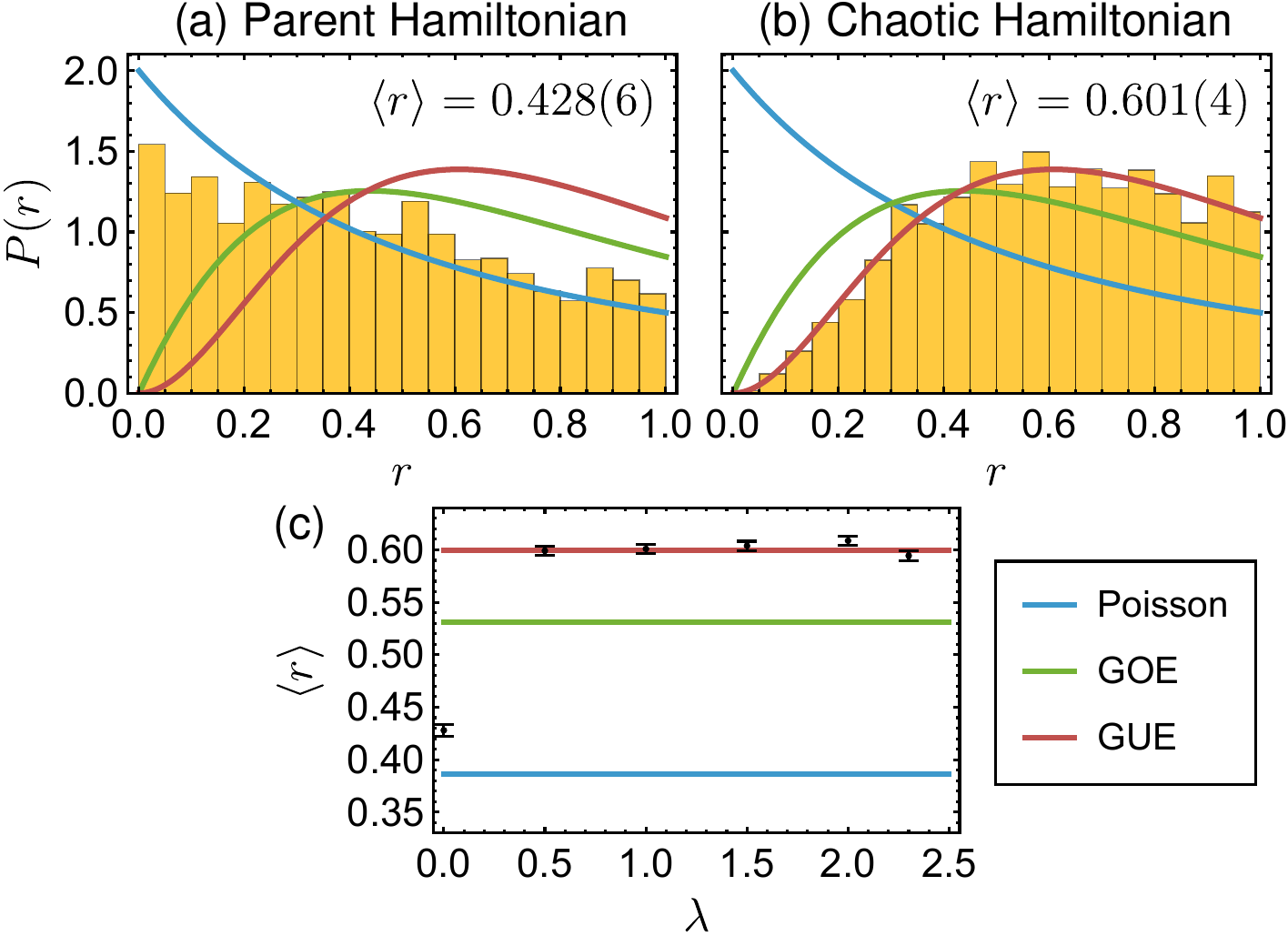}
    \caption{Symmetry-resolved distributions of the ratio $r$ of consecutive level spacings for the generation $g=4$ ($L=17$) Hamiltonians, with spider phases $\phi_g = \pi(1-1/g)$. (a)~The frustration-free parent Hamiltonian~\eqref{eq:parentHam} has near-Poissonian statistics. (b)~After the scar-preserving deformation $\lambda = 1$, the distribution closely follows the GUE prediction. (c)~Average ratio $\langle r \rangle$ as a function of deformation strength $\lambda$. Legend in (c) applies in (a) and (b).}
    \label{fig:statistics}
\end{figure}

We next ask whether the frustration-free Hamiltonian is itself nonintegrable. We diagnose its spectral statistics using the ratio of consecutive level spacings, as described in Sec.~\ref{sec:methods:spectrum}. Spectral statistics are meaningful only after all exact symmetries have been resolved. For the $g=4$, $L=17$ Hamiltonian, we restrict to the sector $\{Z_1=Z_L=+1, R=+1, Q_e=0\}$, and again use the spider phase family of $\phi_g = \pi(1-1/g)$. Because $Q_o$ anticommutes with the two edge operators, it maps this block to the sector $Z_1=Z_L=-1$ and does not further decompose the selected block. However, since both $Q_o$ and $\mathcal A$ flip the edge symmetries, their combined action $Q_o\mathcal A$ is an antiunitary symmetry acting internally within the resolved sector, with $(Q_o\mathcal A)^2 = +1$. Thus, if this block was chaotic, we would expect GOE statistics.

Choosing the specific symmetry block reduces the number of eigenstates enough to be numerically viable.
The resulting distribution $P(r)$ is shown in Fig.~\ref{fig:statistics}(a); it lies closer to the Poisson distribution than to the GOE prediction.
The mean ratio is $\langle r\rangle=0.428(6)$, compared with the Poissonian value $\langle r\rangle_{\text{P}} \approx 0.38629$. The small upward shift can be associated with a slight suppression of small ratios. At the accessible size, the data does not provide evidence that the parent Hamiltonian is chaotic. The statistics is instead consistent with an integrable or near-integrable model, a finite-size crossover, or an additional approximate structure not captured by the exact symmetry resolution.

The parent Hamiltonian features a large zero-energy degeneracy; within the above symmetry sector, the ground-state manifold contains 33 of the 4528 states. Large exact degeneracies of this kind have previously facilitated the identification and analytic construction of exact scar states~\cite{Turner2026, Melendrez2025, Mukherjee2026}. For the sector studied here, we find a spectral gap $\Delta \approx 0.1347$. This gap is comparatively large for the system size $L=17$ and, if persistent, would suggest a gapped thermodynamic limit; however, finite-size scaling is inconclusive with the available system sizes (3, 5, 9, 17, of which the first three likely exhibit large finite-size effects), so the asymptotic fate of $\Delta$ remains uncertain.

We next construct a manifestly quantum-chaotic Hamiltonian that admits the fractal ZX-diagram as an exact, atypical eigenstate.

\subsection{Deformation into the chaotic Hamiltonian}

A local operator that annihilates the target state may be added to the Hamiltonian with an arbitrary coefficient without changing that state or its eigenvalue. This is the essential freedom underlying projector-embedding constructions of exact nonthermal eigenstates~\cite{Shiraishi2017}. However, this often removes most symmetries of the underlying Hamiltonian. In the present setting, ZX-calculus can be used to make the aforementioned deformation local and easy to write down, while also preserving the majority of the symmetry structure.

Specifically, we will use Eq.~\eqref{eq:annihilHb_on_Ha_sites} to define the following perturbation term
\begin{equation}
    V = -\sum_{\substack{i=4,\\i\text{ even}}}^{L-3} H_i^b(g_i),
\end{equation}
which does not change the exact fractal eigenstate when added to the parent Hamiltonian, $H_{\text{par}} + \lambda V$. The deformation destroys the positivity of the full Hamiltonian. As a result, zero energy is no longer constrained to lie at the lower spectral edge; the fractal state is embedded into the bulk of the many-body spectrum.

Specifically, if we set $\lambda=1$, the $H_i^a(k)$ terms on even sites are replaced by $H^{a'}_i(k) = H^a_i(k) - H^b_i(k)$. Interestingly, $H_i^{a'}$ admits a compact ZX representation,
\begin{equation}
    H^{a'}_i(k) =
    \begin{ZX}[column sep={0.6cm,between origins},row sep={0.5cm,between origins}]
        \ar[rr] & & \zxZ{} \ar[rrdd,s] \ar[d] &&\zxN{}\ar[r]& \\
        \ar[r]  & \zxZ*{-\phi_k} & \zxX{\pi} \ar[d]&& \zxZ*{\phi_k} \ar[r]& \\
        \ar[rr] & & \zxZ{} \ar[rruu,s] && \zxN{}\ar[r]&
    \end{ZX}.
\end{equation}
In diagrammatic terms, the difference is simply swapping the outer wires in $H_i^a(k)$~\eqref{eq:parentHam:Ha}; this means that the annihilation in Eq.~\eqref{eq:annihilHa} immediately implies that the $H^{a'}_i(k)$ also annihilates the state.
Importantly, this perturbation preserves every symmetry of the original parent Hamiltonian from Eq.~\eqref{eq:parentHam}, except for the $Q_o$ symmetry; most of the internal symmetry structure is therefore still present. Since $Q_o$ is absent, the remaining antiunitary symmetry $\mathcal A$ does not act internally within the selected symmetry sector $\{Z_1 = Z_L = +1, R = +1, Q_e = 0\}$, so chaotic statistics in this block is expected to be described by GUE.

\begin{figure}
    \centering
    \includegraphics[width=0.89\columnwidth]{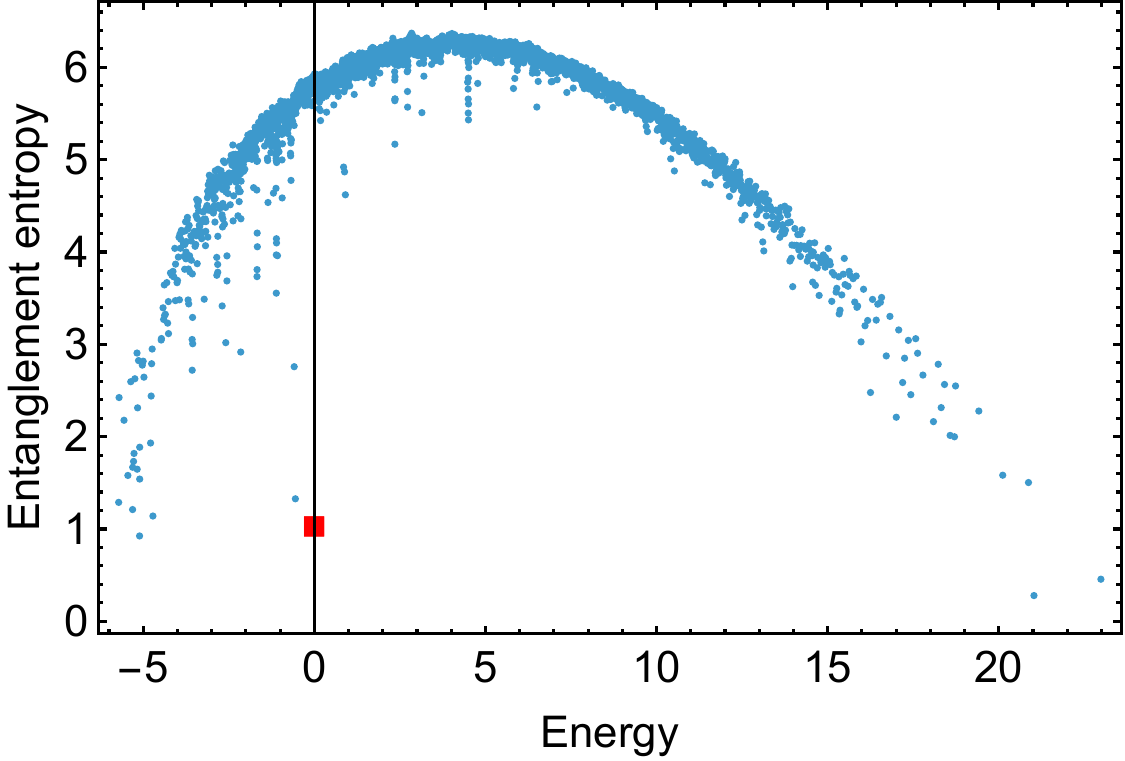}
    \caption{Entanglement entropy vs energy for all nondegenerate eigenstates of the chaotic Hamiltonian for $L=17$ sites. The fractal ZX state is shown by a red square, and can be identified as a quantum many-body scar with area-law entanglement.}
    \label{fig:entvsenergy}
\end{figure}

We find that the ratio of consecutive level gaps for the deformed Hamiltonian closely follows the GUE prediction, as shown in Fig.~\ref{fig:statistics}(b). We obtain $\langle r \rangle = 0.601(4)$, which is completely consistent with the GUE value $\langle r \rangle_{\text{GUE}} \approx 0.5996(1)$. Together with the full distribution $P(r)$, this provides strong finite-size evidence that the deformed Hamiltonian is quantum chaotic within this symmetry sector. In fact, we also find that other non-zero values of $\lambda$ lead to chaotic behavior, as seen in Fig.~\ref{fig:statistics}(c); however, these do not admit simple ZX forms. Adding/subtracting perturbation terms with real coefficients would require extending to ZX-related calculi~\cite{Shaikh2022}.

The triangle fractal ZX-diagram state remains an exact zero-energy eigenstate. Its low entanglement (that could be tuned from area law to log law by changing the phase function $\phi_g$) and atypical local observables, established in Sec.~\ref{sec:fractalZX}, are anomalous relative to the surrounding chaotic spectrum. This is evident from Fig.~\ref{fig:entvsenergy}, where we plot the entanglement entropy vs energy for all nondegenerate eigenstates (the fractal ZX-diagram is marked in red); the state can be readily identified as a quantum many-body scar of the deformed chaotic Hamiltonian. We believe this construction yields a QMBS and an associated chaotic Hamiltonian for any generic $\phi_g$ function, provided that the chosen phases do not introduce additional symmetries that destroy the chaotic character of the spectrum.

In summary, the construction shown here demonstrates a diagrammatic route to Hamiltonian engineering. ZX-calculus insights proved very useful when converting the numerically obtained annihilators into (analytical) local Hamiltonian terms represented as simple ZX-diagrams. A similar diagrammatic reasoning can then be used to perturb from a frustration-free parent model to a chaotic scar-endowed Hamiltonian in a way that preserves most of its symmetries. This suggests that the ZX-calculus can be used as a systematic tool for engineering physically-relevant interacting Hamiltonians with prescribed atypical eigenstates.

\section{Conclusions and outlook}
\label{sec:conclusions}

In this work, we introduced families of fractal many-body states constructed from Sierpi\'nski-triangle and Sierpi\'nski-carpet ZX-diagrams. Both families exhibit anomalously low entanglement and self-similar local observables. The precise entanglement scaling can be controlled by changing the ZX spider phases. Using standard construction methods together with insights from ZX-calculus, we constructed parent Hamiltonians for the triangle ZX states, expressible as compact ZX-diagrams. We then introduced a scar-preserving deformation that moves the state into the spectral bulk and produces Wigner-Dyson level statistics. Thus, the triangle-based fractal ZX-diagrams act as quantum many-body scars for these chaotic Hamiltonians, with tunable entanglement from area law to log law.
These results suggest ZX-calculus as a useful framework for scar construction and Hamiltonian inverse design. The same graphical language describes all the steps: the target scar state construction, identification of the local operators that annihilate it, construction of the frustration-free parent Hamiltonian, and the deformation that renders the surrounding spectrum chaotic. In particular, ZX rewriting exposes the local annihilation mechanism that preserves the exact eigenstate, while being independent of the detailed functional form of the generation-dependent spider phases, allowing for scars with tunable entanglement. 
This implies that ZX‑calculus can be further employed to characterize underlying physics by exposing algebraic structures in many-body quantum states and operators.

Our work opens many interesting future research avenues that combine ZX-calculus and quantum many-body physics. An important direction is to determine how broadly this construction extends beyond the fractal states considered here. The same combination of tensor-network numerics and diagrammatic rewriting could be applied to other exact scar states, matrix-product states, constrained models, and higher-dimensional lattices. The crucial step here was finding annihilation diagrams for local terms, and it would be interesting to investigate whether this can be performed in other models, and systematically. Of particular interest would be the construction of entire scar towers, potentially by identifying diagrammatic counterparts of spectrum-generating algebras or restricted dynamical symmetries. Existing ZX-type calculi can already encode nontrivial lattice-operator algebras~\cite{Coecke2011, East2021, Wang2025}, which could provide diagrammatic methods for proving the required commutator identities for scar towers.

Another question is whether scar-preserving perturbations and Hamiltonian deformation with tunable parameters can also be treated diagrammatically. Representing arbitrary real coefficients and sums of diagrams is not always natural within ordinary ZX-calculus, but the ZXW formalism provides a graphical treatment of linear combinations of scalar-weighted operators~\cite{Shaikh2022}. Moreover, ZXW techniques for exponentiation and Trotterization could be used to study the real-time dynamics of the fractal ZX state directly at the diagrammatic level. More broadly, developing automated methods that combine MPS-based searches for local annihilators~\cite{Ren2025} with ZX or ZXW simplification could provide a systematic route to discovering new local chaotic Hamiltonians with prescribed nonthermal eigenstates.

Finally, the resulting Hamiltonians could be implemented in programmable quantum simulators. Rydberg-atom arrays are a promising setting as blockade constraints naturally generate projector-conditioned dynamics, and recent progress in designing native multibody interactions could help realize the specific three- and five-site terms derived here~\cite{Kerschbaumer2025, Desaules2026, Samajdar2026}. Moreover, the compact ZX representation may also be useful by providing a direct starting point for compiling the local Hamiltonian terms into hardware-native gate sequences implementable on current quantum computing platforms. Taken together, ZX-calculus could help design analytically controlled many-body scars and facilitate their implementation in programmable quantum matter.

The data that support the findings of this article are publicly available~\cite{Data}.

\begin{acknowledgments}
    The author would like to thank Razin Shaikh and Bhaskar Mukherjee for insightful discussions and useful comments. M.~S.\@ was supported by the EPSRC grant on (De)constructing Quantum Software (DeQS) (Grant Reference No. EP/Z002230/1). The author would like to acknowledge the use of the University of Oxford Advanced Research Computing (ARC) facility in carrying out this work~\cite{Richards2015}.
\end{acknowledgments}

\appendix

\section{Supporting data for entanglement entropy and expectation values}
\label{sec:appendix-data}

In this appendix, we show supporting results for the entanglement entropy and the expectation value of $\langle X_i \rangle$ of the fractal ZX-diagrams for phase spider families not shown in the main text.

\begin{figure}[tb]
    \centering
    \includegraphics[width=\columnwidth]{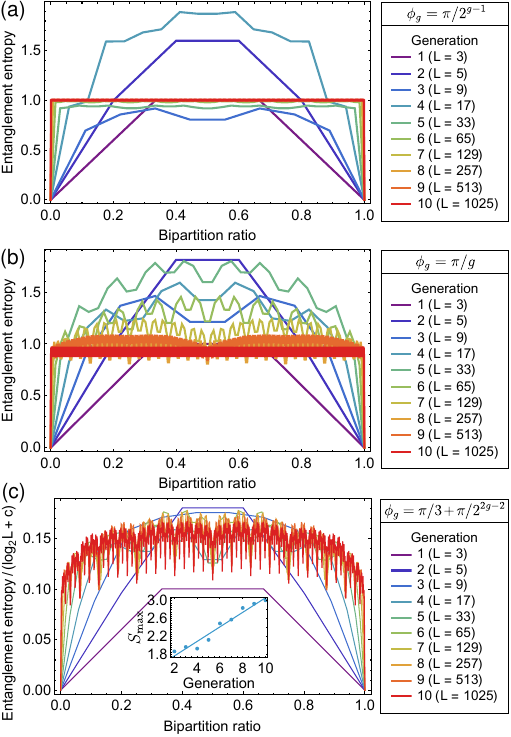}
    \caption{The von Neumann entanglement entropy as a function of the bipartition ratio for successive generations $g$ (and the corresponding system sizes $L$) for the Sierpi\'nski triangle ZX-diagram. Z-spider phases are (a)~$\phi_g = \pi/2^{g-1}$, (b)~$\phi_g = \pi / g$, and (c)~$\phi_g = \pi/3+\pi/2^{2g-2}$ (constant used to rescale the entropy is $c=8$).}
    \label{fig:entanglement_appendix}
\end{figure}

Specifically, Fig.~\ref{fig:entanglement_appendix}(a--b) shows entanglement entropy of the Sierpi\'nski triangle ZX-diagram, with phase spider families $\phi_g = \pi/2^{g-1}$ and $\phi_g = \pi / g$. Both are bounded from above by an $O(1)$ constant, and thus are area-law entangled, similarly to the near-Clifford families presented in the main text. Fig.~\ref{fig:entanglement_appendix}(c) presents the results for $\phi_g=\pi/3+\pi/2^{2g-2}$, which exhibits log scaling of entanglement, similarly to $\phi_g=\pi/3$ from the main text. 

\begin{figure}[!tb]
    \centering
    \includegraphics[width=\columnwidth]{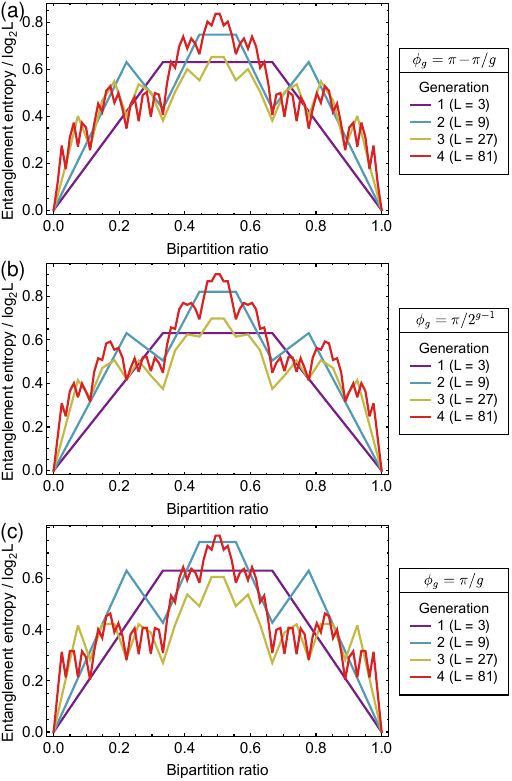}
    \caption{Entanglement entropy divided by $\log_2 L$ as a function of the bipartition ratio for different generations (and the corresponding system sizes) of the Sierpi\'nski carpet ZX-diagram. Z-spider phases are (a)~$\phi_g = \pi(1-1/g)$, (b)~$\phi_g = \pi/2^{g-1}$, and (c)~$\phi_g = \pi/g$.}
    \label{fig:entanglement_carpet_appendix}
\end{figure}

Fig.~\ref{fig:entanglement_carpet_appendix} shows entanglement entropy (rescaled by $\log_2 L$) for the Sierpi\'nski carpet ZX-diagram, with phase spider families $\phi_g = \pi(1-1/g)$, $\phi_g = \pi/2^{g-1}$, and $\phi_g = \pi/g$. Together with the result shown in the main text, all four near-Clifford families show qualitatively the same behavior, with $S/\log_2 L$ seemingly approaching some universal curve. 

\begin{figure}[tb]
    \centering
    \includegraphics[width=\columnwidth]{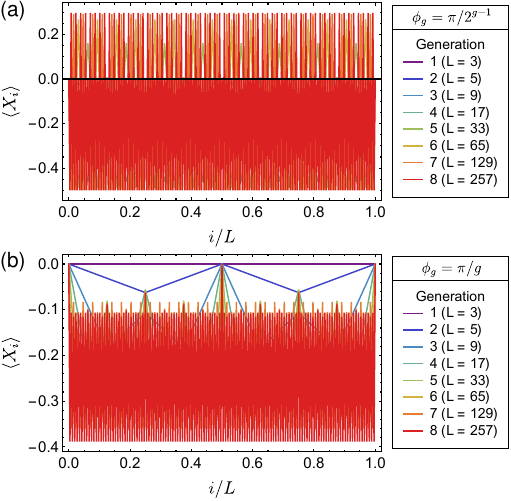}
    \caption{Local expectation value $\langle X_i \rangle$ as a function of the rescaled position $i/L$, for the Sierpi\'nski triangle ZX-diagram. Z-spider phases are (a)~$\phi_g = \pi/2^{g-1}$ and (b)~$\phi_g = \pi/g$.}
    \label{fig:Xexpectation_appendix}
\end{figure}

Fig.~\ref{fig:Xexpectation_appendix} shows the expectation value of $\langle X_i \rangle$ for the Sierpi\'nski triangle with phase spider families $\phi_g = \pi/2^{g-1}$ and $\phi_g = \pi / g$. The results show highly oscillating nonthermal values, similar to the results in the main text.

\section{Example ZX proof of an annihilation identity}
\label{sec:appendix-annihilation}

Here we give a ZX proof of Eq.~\eqref{eq:annihilHa} that certifies annihilation of the state by a local Hamiltonian term~\eqref{eq:parentHam:Ha}. For convenience, we also define two additional rules, which can be derived from the set~\eqref{eq:ZXrules} cited in the main text~\cite{vandeWetering2020}.

The $\pi$-copy rule $(\mathbf{\pi})$ reads
\begin{equation}
    \begin{ZX}[column sep={0.6cm,between origins},row sep={0.6cm,between origins}]
        &&&\zxN{} \ar[r]&\\
        \ar[r]&\zxX{\pi} \ar[r] &\zxZ{\alpha}\ar[r] \ar[ru,)] \ar[rd,(]& \zxN{} \ar[r] &\\
        &&&\zxN{} \ar[r]&
    \end{ZX}
    \overset{(\mathbf{\pi})}{=}
    \begin{ZX}[column sep={0.6cm,between origins},row sep={0.6cm,between origins}]
        &&&\zxX{\pi} \ar[r]&\\
        \ar[r]& \zxN{}\ar[r] &\zxZ*{-\alpha}\ar[r] \ar[ru,)] \ar[rd,(]& \zxX{\pi} \ar[r] &\\
        &&&\zxX{\pi} \ar[r]&
    \end{ZX},
\end{equation}
while the Hopf rule $(\mathbf{hf})$ removes parallel edges between two spiders of different colors (or equivalently, Hadamard parallel edges between two spiders of the same color),
\begin{equation}
    \begin{ZX}[column sep={0.6cm,between origins},row sep={0.3cm,between origins}]
        \ar[r]&\zxZ{} \ar[rr,)] \ar[rr,(] &&\zxX{} \ar[r]&
    \end{ZX}
    \overset{(\mathbf{hf})}{=}
    \begin{ZX}[column sep={0.6cm,between origins},row sep={0.3cm,between origins}]
        \ar[r]&\zxZ{} &&\zxX{} \ar[r]&
    \end{ZX}.
\end{equation}
Note the latter also implies that simple-edge self-loops can be removed, while Hadamard-edge self-loops can be removed after adding a $\pi$ phase to the spider.

The annihilation identity can then be proven by simplifying the left side to a zero diagram,
\begin{widetext}
\begin{subequations}
\begin{align}
    &\begin{ZX}[column sep={0.3cm,between origins},row sep={0.6cm,between origins}]
        &&&\ar[dr]&&&&&&&&&&&&\\
        &&\ar[rr]&&\zxZ*{\alpha} \ar[dr,blue,dashed] \ar[dl,blue,dashed] \ar[rrrrrrr] &&&&&&& \zxZ{} \ar[rrrr] \ar[d] &&&&\\
        &&& \zxZ*{\phi_k} \ar[dl,blue,dashed] \ar[dr,blue,dashed] \ar[rr,blue,dashed] && \zxZ*{\phi_k} \ar[dl,blue,dashed] \ar[dr,blue,dashed] \ar[rrrr] &&&&\zxZ*{-\phi_k}&&\zxX{\pi}\ar[d]&&\zxZ*{\phi_k} \ar[rr]&&\\
        \ar[rr]&&\zxZ*{\beta}\ar[rr,blue,dashed] \ar[dl]&&\zxZ*{\phi_k}\ar[rr,blue,dashed]&&\zxZ*{\gamma} \ar[rrrrr] \ar[dl] \ar[dr]&&&&&\zxZ{}\ar[rrrr]&&&&\\
        &&&&&&&&&&&&&&&
    \end{ZX}
    \overset{(\mathbf{f})}{=}
    \begin{ZX}[column sep={0.3cm,between origins},row sep={0.6cm,between origins}]
        &&&\ar[dr]&&&&&&&&\\
        &&\ar[rr]&&\zxZ*{\alpha} \ar[dr,blue,dashed] \ar[dl,blue,dashed] \ar[rrrrrrr] \ar[drrr] &&&&&&&\\
        &&& \zxZ*{\phi_k} \ar[dl,blue,dashed] \ar[dr,blue,dashed] \ar[rr,blue,dashed] && \zxZ*{} \ar[dl,blue,dashed] \ar[dr,blue,dashed] &&\zxX{\pi}\ar[dl]&&\zxZ*{\phi_k} \ar[rr]&&\\
        \ar[rr]&&\zxZ*{\beta}\ar[rr,blue,dashed] \ar[dl]&&\zxZ*{\phi_k}\ar[rr,blue,dashed]&&\zxZ*{\gamma} \ar[rrrrr] \ar[dl] \ar[dr]&&&&&\\
        &&&&&&&&&&&
    \end{ZX}
    \overset{(\mathbf{\pi})}{\overset{(\mathbf{f})}{=}}
    \begin{ZX}[column sep={0.35cm,between origins},row sep={0.6cm,between origins}]
        &&&\ar[dr]&&&&&&&&\\
        &&\ar[rr]&&\zxZ*{\alpha} \ar[dr,blue,dashed] \ar[dl,blue,dashed] \ar[rrrrrrr] \ar[ddrr,)] &&&&&&&\\
        &&& \zxZ*{\phi_k} \ar[dl,blue,dashed] \ar[dr,blue,dashed] \ar[rr,blue,dashed] && \zxZ*{\pi} \ar[dl,blue,dashed] \ar[dr,blue,dashed] &&&&\zxZ*{\phi_k} \ar[rr]&&\\
        \ar[rr]&&\zxZ*{\beta}\ar[rr,blue,dashed] \ar[dl]&&\zxZ*{\phi_k+\pi}\ar[rr,blue,dashed]&&\zxZ*{-\gamma} \ar[rr] \ar[dl] \ar[dr]&&\zxX{\pi} \ar[rrr]&&&\\
        &&&&&\zxX{\pi} \ar[dl]&&\zxX{\pi} \ar[dr]&&&&\\
        &&&&&&&&&&&
    \end{ZX}\\
    & \qquad \overset{(\mathbf{f})}{=}
    \begin{ZX}[column sep={0.35cm,between origins},row sep={0.6cm,between origins}]
        &&&\ar[drrrrr]&&&&&&&&\\
        &&\ar[rrrrrr]&&&&&&\zxZ*{\alpha-\gamma} \ar[rrr] \ar[dlll,blue,dashed,'>] \ar[dlll,blue,dashed,-o={a=60}] \ar[dlllll,blue,dashed,'>] \ar[dd] \ar[ddllll,blue,dashed,-o={a=45}] \ar[dddl] \ar[dddlll]&&&\\
        &&& \zxZ*{\phi_k} \ar[dl,blue,dashed] \ar[dr,blue,dashed] \ar[rr,blue,dashed] && \zxZ*{\pi} \ar[dl,blue,dashed] &&&&\zxZ*{\phi_k} \ar[rr]&&\\
        \ar[rr]&&\zxZ*{\beta}\ar[rr,blue,dashed] \ar[dl]&&\zxZ*{\phi_k+\pi} &&&&\zxX{\pi} \ar[rrr]&&&\\
        &&&&&\zxX{\pi} \ar[dl]&&\zxX{\pi} \ar[dr]&&&&\\
        &&&&&&&&&&&
    \end{ZX}
    \overset{(\mathbf{hf})}{=}
    \begin{ZX}[column sep={0.35cm,between origins},row sep={0.6cm,between origins}]
        &&&\ar[drrrrr]&&&&&&&&\\
        &&\ar[rrrrrr]&&&&&&\zxZ*{\alpha-\gamma} \ar[rrr] \ar[dlllll,blue,dashed,'>] \ar[dd] \ar[ddllll,blue,dashed,-o={a=45}] \ar[dddl] \ar[dddlll]&&&\\
        &&& \zxZ*{\phi_k} \ar[dl,blue,dashed] \ar[dr,blue,dashed] \ar[rr,blue,dashed] && \zxZ*{\pi} \ar[dl,blue,dashed] &&&&\zxZ*{\phi_k} \ar[rr]&&\\
        \ar[rr]&&\zxZ*{\beta}\ar[rr,blue,dashed] \ar[dl]&&\zxZ*{\phi_k+\pi} &&&&\zxX{\pi} \ar[rrr]&&&\\
        &&&&&\zxX{\pi} \ar[dl]&&\zxX{\pi} \ar[dr]&&&&\\
        &&&&&&&&&&&
    \end{ZX}
    \overset{(\mathbf{\pi})}{\overset{(\mathbf{f})}{=}}
    \begin{ZX}[column sep={0.35cm,between origins},row sep={0.6cm,between origins}]
        &&&\ar[drrrrr]&&&&&&&&\\
        &&\ar[rrrrrr]&&&&&&\zxZ*{\alpha-\gamma+\pi} \ar[rrr] \ar[dlllll,blue,dashed,'>] \ar[dd] \ar[ddllll,blue,dashed,-o={a=45}] \ar[dddl] \ar[dddlll]&&&\\
        &&& \zxZ*{-\phi_k} \ar[dl,blue,dashed] \ar[dr,blue,dashed] \ar[dr,)] &&&&&&\zxZ*{\phi_k} \ar[rr]&&\\
        \ar[rr]&&\zxZ*{\beta+\pi}\ar[rr,blue,dashed] \ar[dl]&&\zxZ*{\phi_k} &&&&\zxX{\pi} \ar[rrr]&&&\\
        &&&&&\zxX{\pi} \ar[dl]&&\zxX{\pi} \ar[dr]&&&&\\
        &&&&&&&&&&&
    \end{ZX}\\
    &\qquad\overset{(\mathbf{f})}{=}
    \begin{ZX}[column sep={0.35cm,between origins},row sep={0.6cm,between origins}]
        &&&\ar[drrrrr]&&&&&&&&\\
        &&\ar[rrrrrr]&&&&&&\zxZ*{\alpha-\gamma+\pi} \ar[rrr] \ar[dllll,blue,dashed,'>] \ar[dd] \ar[dllll,blue,dashed,-o={a=45}] \ar[dddl] \ar[dddlll]&&&\\
        &&&& \zxZ*{} \ar[dll,blue,dashed,(] \ar[dll,blue,dashed,)] \zxLoop[120][40][blue,dashed]{} &&&&&\zxZ*{\phi_k} \ar[rr]&&\\
        \ar[rr]&&\zxZ*{\beta+\pi} \ar[dl]&&&&&&\zxX{\pi} \ar[rrr]&&&\\
        &&&&&\zxX{\pi} \ar[dl]&&\zxX{\pi} \ar[dr]&&&&\\
        &&&&&&&&&&&
    \end{ZX}
    \overset{(\mathbf{hf})}{=}
    \begin{ZX}[column sep={0.35cm,between origins},row sep={0.6cm,between origins}]
        &&&\ar[drrrrr]&&&&&&&&\\
        &&\ar[rrrrrr]&&&&&&\zxZ*{\alpha-\gamma+\pi} \ar[rrr] \ar[dd] \ar[dddl] \ar[dddlll]&&&\\
        &&&& \zxZ*{\pi} &&&&&\zxZ*{\phi_k} \ar[rr]&&\\
        \ar[rr]&&\zxZ*{\beta+\pi} \ar[dl]&&&&&&\zxX{\pi} \ar[rrr]&&&\\
        &&&&&\zxX{\pi} \ar[dl]&&\zxX{\pi} \ar[dr]&&&&\\
        &&&&&&&&&&&
    \end{ZX} = 0,
\end{align}
\end{subequations}
\end{widetext}
where the last equality comes from the standalone spider with $\pi$ phase being zero, $\begin{ZX}
    \zxZ{\pi}
\end{ZX}=1+e^{i\pi}=0$. This annihilation identity and the other diagrammatic identities present in the main text can be verified using similar simplification techniques as well as the automatic simplification methods in PyZX~\cite{Kissinger2020PyZX} and ZXLive~\cite{ZXLive}.

\bibliography{refs}

\end{document}